\documentclass{jfm}

\usepackage{graphicx}
\usepackage{newtxtext}
\usepackage{float}
\usepackage{newtxmath}
\usepackage{natbib}
\usepackage{hyperref}
\usepackage{cleveref}
\usepackage{comment}
\hypersetup{
    colorlinks = true,
    urlcolor   = blue,
    citecolor  = black,
}

\newcommand{\RomanNumeralCaps}[1]

\usepackage{xspace}
\newcommand{\St}{\ensuremath{\mathrm{\textit{St}}}\xspace}

\title{Fluid memory beyond creeping flow: a history kernel for a sphere at finite Reynolds numbers}

\author{M. D. Łasica\aff{1}  \corresp{\email{m.d.lasica@tue.nl}},
        J. G. M. Kuerten\aff{1}
        \and 
        S. K. F. Stoter\aff{1}}

\affiliation{\aff{1}Eindhoven University of Technology, Department of Mechanical Engineering, P.O. Box 513,
5600 MB, Eindhoven, The Netherlands}

\begin{document}
\raggedbottom
\maketitle

\begin{abstract}
Predicting the history force on a sphere beyond the creeping-flow limit remains challenging. Existing formulations have been derived or calibrated for selected flow histories, with limited evidence of their predictive accuracy under more general conditions. We develop a systematic approach to extract the history kernel of a fixed sphere in spatially uniform, unsteady flow from axisymmetric Navier-Stokes simulations of finite-duration acceleration and deceleration ramps at initial Reynolds numbers from 0.2 to 200. The resulting kernel surface recovers the Basset limit at short lags and in the creeping-flow limit and is continued beyond the resolved lags using far-wake asymptotics. At higher Reynolds numbers, the kernel develops non-monotonic behaviour at intermediate lags, associated with delayed adjustment of the boundary layer and wake. For the cases tested, the present kernel substantially improves history-force predictions for finite ramps, oscillatory forcing and non-periodic composite velocity histories relative to the literature formulations considered, without case-specific adjustment. These findings indicate that lag and Reynolds number capture the dominant memory response. 
\end{abstract}

\begin{keywords}
history force, unsteady flow past a sphere, particle–fluid interactions
\end{keywords}

\section{Introduction}
\label{sec:headings}
Particles suspended in flows are ubiquitous in natural and engineering systems. Euler-Lagrange point-particle methods provide an efficient way to track large populations of such particles without resolving the flow around each one. Their accuracy therefore depends on hydrodynamic force models that represent unresolved particle-fluid interactions. Predicting particle motion requires accounting for both quasi-steady and unsteady forces arising from the relative motion between the particle and fluid. Among these forces, the history force remains poorly understood beyond the creeping-flow limit, making its finite-\(\Rey\) behaviour a key challenge for reliable particle tracking.

The equation of motion for a finite-size spherical particle moving through an unsteady, spatially non-uniform Stokes flow was formulated by \citet{MR1983}. Neglecting the Faxén corrections, this equation can be written as
\begin{equation}
\begin{aligned}
m_{\mathrm p}\,
\frac{\mathrm{d}\boldsymbol{v}}{\mathrm{d}t}
={}&
m_{\mathrm f}\,
\frac{\mathrm{D}\boldsymbol{u}}{\mathrm{D}t}
+\frac{1}{2}\,m_{\mathrm f}\,
\left(
\frac{\mathrm D\boldsymbol{u}}{\mathrm Dt}
-
\frac{\mathrm d\boldsymbol{v}}{\mathrm dt}
\right) + 6\,r\,\upi\,\mu\,
\left(\boldsymbol{u}-\boldsymbol{v}\right)
\\
& +6\,r^{2}\sqrt{\upi\,\rho_{\mathrm f}\,\mu}\,
\int_{0}^{t}
\frac{\mathrm{d}\big(\boldsymbol{u}(\tau)-\boldsymbol{v}(\tau)\big)}
     {\mathrm{d}\tau}\,
\frac{1}{\sqrt{(t-\tau)}}\,\mathrm{d}\tau
+\left(m_{\mathrm p}-m_{\mathrm f}\right)\,\boldsymbol{g} \,.
\end{aligned}
\label{eq:particle_motion}
\end{equation}
Here \(m_{\mathrm p}\) is the particle mass and \(m_{\mathrm f}=4/3\,\upi\,r^{3}\,\rho_{\mathrm f}\) is the mass of the displaced fluid, with \(\rho_{\mathrm f}\) representing the fluid density. The particle velocity is denoted by
\(\boldsymbol{v}\), while \(\boldsymbol{u}\) is the undisturbed-fluid velocity evaluated at the particle centre. The operator $\mathrm{D}/\mathrm{D}t$ denotes the material derivative following the undisturbed fluid, whereas $\mathrm{d}/\mathrm{d}t$ denotes the time derivative along the particle trajectory. Furthermore, \(r\) is the particle radius, \(\mu\) is the dynamic viscosity and \(\boldsymbol{g}\) is the gravitational acceleration. The time variable \(t\) denotes the dimensional current time. 

The Maxey-Riley equation is often regarded as a generalisation of the Boussinesq-Basset-Oseen equation. The left-hand side of \cref{eq:particle_motion} represents particle inertia. The terms on the right-hand side describe, respectively, the undisturbed-flow (or pressure-gradient) force, the added mass force, the quasi-steady Stokes drag, the history force and the combined gravitational and buoyancy force. The history force represents the viscous diffusion of vorticity generated by past changes in the particle-fluid relative velocity, thereby introducing a dependence on the particle's previous relative acceleration. The Maxey-Riley equation assumes that advective inertia in the disturbance flow is negligible. At finite Reynolds numbers, however, this assumption breaks down: convective transport and wake development modify both the quasi-steady drag and the temporal memory represented by the history force. Although finite-\(\Rey\) corrections to quasi-steady drag are well established, corresponding formulations for the history force remain comparatively scarce.
\subsection{A review of history force} 
\label{sec:history_force_rev}
The history force was first derived from the unsteady Stokes equations by
\citet{BOUSSINESQ1885} and, independently three years later, by
\citet{Basset1888}. To simplify the notation, we define
\begin{equation}
\boldsymbol{F}_{\mathrm H}(t)
=
\Phi\,
\int_{0}^{t}
\boldsymbol{\alpha}(\tau)\,
K_{\mathrm B}(t-\tau)\,\mathrm{d}\tau \,,
\end{equation}
with
\begin{equation}
\Phi=6\,r\,\upi\,\mu,
\quad
\boldsymbol{\alpha}(\tau)
=
\frac{\mathrm{d}\big(
\boldsymbol{u}(\tau)-\boldsymbol{v}(\tau)
\big)}{\mathrm{d}\tau} \,,
\quad
K_{\mathrm B}(t-\tau)
=
\frac{r}
{\sqrt{\upi\,\nu_f\,(t-\tau)}} \,.
\end{equation}
Here \(\nu_f=\mu /\rho_f\) is the kinematic viscosity. The history force is a convolution of the relative acceleration with a history kernel that weighs each contribution according to its lag \(t-\tau\). The kernel decays algebraically as \(t^{-1/2}\), therefore recent changes in slip velocity contribute more strongly than older ones. For consistency with the literature, we refer to \(K_B\) as the Basset kernel. 

An early attempt to extend the Basset force beyond the Stokes regime was made by \citet{Odar_Hamilton_1964}, who measured the force on a sphere in rectilinear harmonic motion in a quiescent fluid at Reynolds numbers \(\Rey=2\rho_f ur/\mu\) up to 64 and acceleration ratios up to 4. The paper demonstrates a particularly clear fluid memory effect: after one oscillation the sphere can return to the same position, velocity and acceleration, but experiences a different force because the liquid has retained motion from the preceding cycle. Since these hysteresis results confirmed the expected deviation from the Boussinesq-Basset-Oseen equations, the authors introduced empirical coefficients  \(C_\mathrm{V}\), \(C_\mathrm{A}\) and \(C_\mathrm{H}\), multiplying the quasi-steady, added-mass, and history force terms, respectively. The latter two were formulated as functions of the acceleration number alone.

\citet{Mei_Adrian_1992} later argued that multiplying the  Basset integral by an acceleration-dependent coefficient is not physically justified, because finite-\(\Rey\) effects change the kernel shape, not only its amplitude. They further attributed the apparent agreement with experimental data to a cancellation between the history and added-mass contributions. The added-mass coefficient should therefore be retained at its theoretical value of \(1/2\) \citep{Batchelor_1967}, consistent with the numerical findings of \citet{Mei_Lawrence_Adrian_1991} and \citet{Chang_Maxey_1995}. Subsequently, \citet{Sano_1981} considered an impulsive change in relative velocity from zero to a constant value at small but non-zero \(\Rey\). These results showed that the transient composite drag expression retains the Basset \(t^{-1/2}\) behaviour at short times but decays as \(t^{-2}\) at long times because free-stream advection cancels the leading diffusive contribution. The analysis revises the matched-asymptotic treatment of \citet{Bentwich_Miloh_1978} through a three-region structure. Although the expansion fails quantitatively against the full Navier-Stokes solution of \citet{Dennis_Walker_1972} at moderate \(\Rey\), it establishes the essential mechanism: advection of vorticity by the free stream removes the fluid's long-time memory of the particle motion. 

\citet{Mei_Lawrence_Adrian_1991} numerically computed the unsteady drag on a stationary sphere at Reynolds numbers between 0.1 and 40, subject to small harmonic fluctuations of the free-stream velocity. The harmonic force was decomposed into the quasi-steady response, the contributions of the undisturbed-flow and added-mass forces, and a remaining acceleration-dependent residual identified as the modified Basset force. With \(\omega\) denoting the angular frequency, this residual disagrees with the classical prediction at low frequency: instead of the \(\sqrt{\omega}\) scaling of the unsteady Stokes solution, it vanishes linearly with  \(\omega\), recovering the classical behaviour only at high frequency. Since the long-time behaviour of the memory kernel is dictated by the low-frequency limit of the force, this implies that the kernel retains the Basset \(t^{-1/2}\) form at short times but decays considerably faster at long times. Regardless of flow conditions, the computed history force decreases monotonically with \(\Rey\), always remaining below the unsteady-Stokes solution. The classical Basset force therefore systematically overestimates the finite-\(\Rey\) history force.

In a follow-up study, \citet{Mei_Adrian_1992} used matched asymptotic expansions in the limit \(\St\ll \Rey\ll1\), where \(\St=\omega\,r/u\) denotes the Strouhal number. They showed that the acceleration-dependent drag (pressure gradient force, added mass and history force) at low frequency is exactly \(0.75\,\St\) at leading order, which confirmed the linear frequency dependence previously observed numerically. The authors extended the low-\(\Rey\) result using a curve fit to the available finite-difference calculations. They then combined three sources of information: the analytical low-frequency limit, the unsteady-Stokes solution at high frequency, and the numerical results at intermediate frequencies. From these limits, they could interpolate the history-force response in the frequency domain, yielding the following expression: 
\begin{equation}
K_\mathrm{MA}(t-\tau,\Rey) 
=
\left[
K_{\mathrm B}(t-\tau)^{-1/c_1}
+
\left(
\frac{2\,r\,\nu_f\,f_{\mathrm H}^{3}(\Rey)}
     {\upi\,u^{3}\,(t-\tau)^{2}}
\right)^{-1/c_1}
\right]^{-c_1},
\qquad
\label{eq:MA_kernel}
\end{equation}
with 
\begin{equation}
   f_{\mathrm H}(\Rey)=0.75+c_2\,\Rey
\end{equation}
Here, \(c_\mathrm{1}=2\) controls the shape and sharpness of the harmonic mean-type interpolation between the short- and long-time asymptotic limits, whereas \(c_\mathrm{2}=0.105\) is the empirical finite-\(\Rey\) correction to the low-frequency history force coefficient. The kernel recovers \((t-\tau)^{-1/2}\) at short lag and transitions to \((t-\tau)^{-2}\) at long lag, for both small and finite-\(\Rey\). The short-lag limit is well established, but the detailed form of the \(t^{-2}\) tail is partly imposed by the interpolation and was not uniquely fixed by the available frequency domain asymptotes.

\citet{Mei_1993} tested the kernel for a step change in free-stream velocity using time-domain Navier-Stokes calculations for \(0.1\leq\Rey\leq100\). The impulsive acceleration isolates the history force after the initial instant. The response followed \(t^{-1/2}\) at short times and \(t^{-n}\) at intermediate times, with \(n\approx2\) for \(\Rey<10\) and slightly larger for  higher \(\Rey\), before an apparently exponential regime was reported for dimensionless time \(t'>100\). Comparable decay was obtained for start-up from quiescent fluid and acceleration from a developed steady wake, suggesting limited sensitivity to the initial vorticity field, although only two initial conditions cannot establish complete independence. For zero-mean oscillatory motion with flow reversal, \citet{Mei_1994} found an \(St\,\log St\) contribution to the low-frequency imaginary component of the drag amplitude rather than the \(\mathcal{O}(St)\) dependence of \citet{Mei_Adrian_1992}, which is recovered when strong mean convection suppresses the logarithmic term. The kernel was not modified because the  postulated coefficient depends on the ratio of oscillation amplitude to mean flow, which is not readily defined for a particle undergoing general unsteady motion. Moreover, the resulting discrepancy was modest and all forms agreed at high frequency. \citet{Mei_1994} nevertheless argued that the kernel should be evaluated using the velocity and Reynolds number at the integration time rather than at the current time. 

A complementary numerical study investigating a fixed sphere in a zero-mean sinusoidally oscillating uniform flow at \(\Rey\lesssim20\) was provided by \citet{Chang_Maxey_1994}. The unsteady-Stokes response was recovered at low \(\Rey\) and small amplitude, whereas at finite amplitude flow separation occurred during deceleration at Reynolds numbers for which the corresponding steady flow remained attached. The separated vorticity persisted into the next acceleration phase, so equal instantaneous velocities could produce different forces. The authors therefore avoided decomposing the force into quasi-steady and history contributions, because the vorticity field was generally different from the steady flow at the same instantaneous Reynolds number. The study demonstrates that finite-\(\Rey\) memory depends not only on the oscillation amplitude, but also on phase and state of the wake. Their later calculations at \(\Rey\leq10\) showed stronger wake modification during deceleration, including complete vorticity-layer detachment and negative viscous force while the free-stream velocity remained positive \citep{Chang_Maxey_1995}. The initial relaxation remained approximately proportional to \(t^{-1/2}\). At later times, non-zero final velocities produced an apparently exponential stage, whereas complete stopping gave algebraic decay.

The ultimate long-time behaviour was clarified by \citet{Lawrence_Mei_1995}, who combined far-wake analysis \citep{Batchelor_1967} with numerical solutions of the Navier-Stokes equations up to \(\Rey\) of 60 for abrupt changes between steady velocities. At large times, the flow consists of an old wake associated with the initial velocity and a new wake associated with the final velocity, connected by a diffusing transition region convected downstream at the new free-stream speed. The source or sink required by mass conservation in this region controls the transient force. This gives \(t^{-2}\) relaxation for start-up, changes between two positive velocities and non-parallel changes of direction, but \(t^{-1}\) for complete stopping or exact reversal because the body remains within, or returns through, its earlier wake. Furthermore, calculations extending to dimensionless times of about 300 attributed the exponential regime of \citet{Mei_1993} to insufficient far-wake resolution. The study thus determines the asymptotic decay rates but does not provide a complete history kernel connecting the short-time Basset response to the far-wake regime. This is evident in the intermediate region, where the force departs from the \(t^{-1/2}\) scaling before reaching the  \(t^{-2}\) asymptote. \citet{Lovalenti_Brady_1995} incorporated this far-wake correction into a small-\(\Rey\) unsteady-Oseen theory for arbitrary motion, in which memory depends on both lag and the displacement variable
\begin{equation}
\boldsymbol{A}(t,\tau)
\propto
\frac{1}{\sqrt{t-\tau}}
\,
\int_{\tau}^{t}
\boldsymbol{U}_{\mathrm{s}}(q)\,\mathrm{d}q,
\end{equation}
where \(\boldsymbol{U}_{\mathrm s}(q)\) is the particle-fluid slip velocity. The variable \(\boldsymbol{A}(t,\tau)\) compares the accumulated relative displacement with the corresponding viscous diffusion length.  The model distinguishes start-up, stopping, reversal and transitions between non-zero velocities, recovering the corresponding \(t^{-2}\) and \(t^{-1}\) limits. It therefore describes finite-\(\Rey\) memory as a nonlinear, path-dependent functional rather than a lag-only convolution. Since convection is represented by the linearised Oseen approximation and the formulation was tested only for \(\Rey\leq1\), it extends temporal validity rather than the Reynolds number range.

Later models sought broader practical applicability.
\citet{KIM_ELGHOBASHI_1998} compared the Boussinesq-Basset-Oseen,
Odar-Hamilton and Mei-Adrian formulations with axisymmetric Navier-Stokes
simulations of freely moving spheres for \(2\leq\Rey\leq150\). Mei-Adrian
performed best overall but became less accurate during strong deceleration.
Their modification used the instantaneous acceleration parameters and approached the Basset and Mei-Adrian forms at large and small acceleration number, respectively. However, its six empirical constants were fitted to the same simulations used for assessment, so the reported accuracy is not an independent validation. \citet{WAKABA_2005} provided a direct time-domain kernel extraction for acceleration from \(\Rey=50\) to \(62.5\) and the reverse deceleration. The accelerating kernel retained the \(t^{-1/2}\) short-lag limit, then developed a pronounced hump before decaying faster than \(t^{-2}\). The hump became more pronounced for acceleration from \(\Rey=100\) to \(125\). Acceleration and deceleration produced different kernels beyond the short-lag regime, despite traversing the same Reynolds number interval in opposite directions. This asymmetry was attributed to the different pre-existing wakes and showed that the response may depend on the path between the initial and final states. Finally, \citet{DORGAN2007} recalibrated the Mei-Adrian kernel to \(c_1=2.5\) and \(c_2=0.2\). The calibration relied mainly on the settling experiments of \citet{Moorman1955}, covering terminal Reynolds numbers from 9 to 853. It improved their trajectory predictions relative to the original Mei-Adrian and approximate Kim models, which generally underestimated the history force. Agreement with the forced-oscillation experiments of \citet{Odar_Hamilton_1964} was weaker, particularly during rapid deceleration and reversal, which also cautions against assuming a universal \(t^{-2}\) tail. Since the Moorman data were also used for calibration, their agreement is not an independent validation. The Dorgan-Loth kernel should therefore be viewed as a calibrated model for relatively simple finite-\(\Rey\) motions, with limited evidence for arbitrary trajectories.
\subsection{Research gap and contributions of the present work} 
\label{sec:research_gap}
\begin{figure}
    \centering
    \includegraphics[width=1\linewidth]{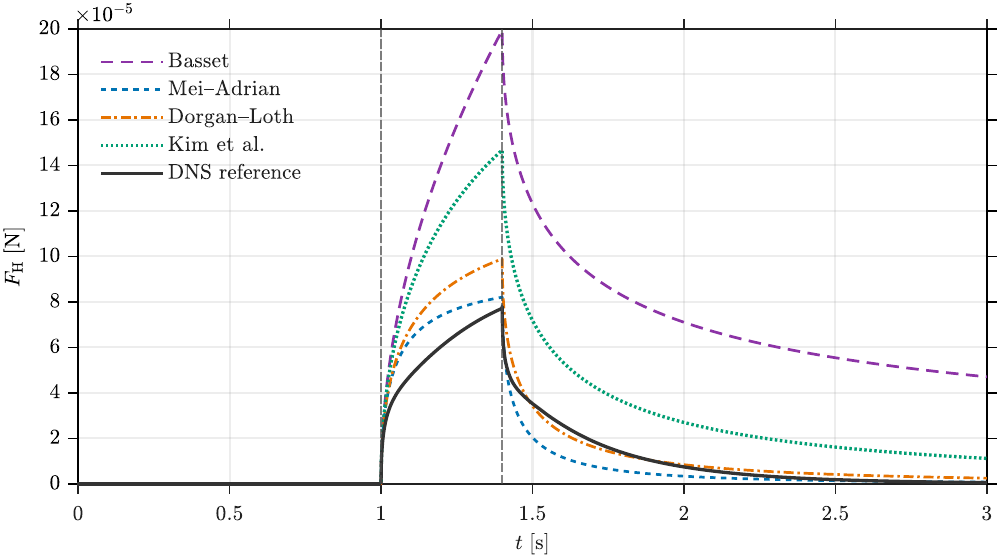}
    \caption{Comparison of the numerical history force with the
    classical Basset, Mei-Adrian, Dorgan-Loth and Kim formulations
    for a linear velocity ramp from \(\Rey=90\) to
    \(\Rey=99\) over \(\Delta t_r=0.4\,\mathrm{s}\). Dashed vertical lines indicate the start and end of the acceleration window. All formulations overpredict the peak force and differ from the numerical post-ramp decay.}
    \label{fig:history_force_benchmark}
\end{figure}
The preceding review reveals a more fundamental limitation than uncertainty in model coefficients. Existing history-force formulations were derived, calibrated or assessed for restricted combinations of Reynolds number, forcing frequency, acceleration amplitude and preceding flow history. Therefore, the reported Reynolds number ranges do not necessarily indicate the full ranges over which these models have been independently shown to give accurate predictions. Agreement with calibration data establishes the quality of the fit to those data, but does not establish predictive accuracy for independent flow histories. The present gap is therefore not simply the absence of another finite-\(\Rey\) correction, but the absence of a history kernel obtained using a common extraction protocol and demonstrated to remain predictive over a broad Reynolds number interval without case-specific retuning. 

To illustrate the gap, the classical Basset, Mei-Adrian, Dorgan-Loth and Kim formulations were evaluated using the same prescribed velocity history, force decomposition and numerical reference. The finite-\(\Rey\) kernels were evaluated at the source-time Reynolds number. A representative ramp from \(\Rey=90\) to \(\Rey=99\), lying within the reported ranges of the finite-\(\Rey\) formulations considered, is shown in \cref{fig:history_force_benchmark}. All four formulations overpredict the maximum history force. The Basset and Kim models retain excessive long-time memory, whereas the Mei-Adrian response decays too rapidly. The Dorgan-Loth formulation provides the closest agreement in the decay phase, but the Mei-Adrian model better captures the acceleration phase. Across the range of Reynolds numbers considered, the numerical tail above \(\Rey\approx40\) also develops a different functional form that cannot be reproduced by changing only the coefficients of a Mei-Adrian type kernel.

The present work addresses this gap using a finite-duration ramp approach that recovers the history kernel without prescribing its analytical form over the lag range resolved. Simulations of both accelerating and decelerating flows are used to identify the kernel \(K(s,\Rey(t_0))\) over \(0.2\leq\Rey\leq200\). The kernels obtained at individual Reynolds numbers are combined into a surface that preserves the classical Basset limit as both \((t-\tau)\rightarrow0\) and \(\Rey \rightarrow0\). Finally, the surface is used to compute the history force for different velocity histories, and the resulting predictions are benchmarked against literature formulations and numerical solutions under identical conditions.

The paper is organised as follows. Section 2 presents the kernel construction methodology, while Section 3 describes the numerical simulations and the generated database. Section 4 examines kernel structure and the associated wake dynamics, and Section 5 assesses predictive performance. Section 6 discusses the findings and presents the conclusions.

\section{Kernel construction methodology}
\subsection{Ramp-flow and lag-window formulation}
We consider a rigid spherical particle held fixed in an incompressible Newtonian fluid. The imposed undisturbed flow is spatially uniform but unsteady. Under these assumptions, and neglecting gravity, we adopt the finite-\(\Rey\) generalisation of \cref{eq:particle_motion}, which reads:
\begin{equation}
\label{eq:fixed_sphere_force}
F(t) =  \frac{3}{2}\,m_{\mathrm f}\,
 \frac{\mathrm d u_\infty}{\mathrm dt} + F_\mathrm{QS}(u_\infty(t)) +  F_\mathrm{H}(t).
\end{equation}
The terms on the right-hand side represent, respectively, the combined pressure-gradient and added-mass contribution, the quasi-steady drag  and the history force. The first term has an exact closed-form expression and the second is evaluated using the steady-state force at the instantaneous free-stream velocity \(u_\infty\). Thus, the remaining unsteady contribution defines~\(F_{\mathrm H}\). 

To initiate the unsteady response in a controlled manner, we prescribe a finite duration constant acceleration ramp,
\begin{equation}
u_\infty(t)
=
\begin{cases}
u_0,
    & t<t_0, \\[2mm]
u_0+\alpha\,(t-t_0),
    & t_0\leq t\leq t_0+\Delta t_r, \\[2mm]
u_0+\Delta u,
    & t>t_0+\Delta t_r,
\end{cases}
\qquad
\alpha=\frac{\Delta u}{\Delta t_r}.
\label{eq:ramp_velocity}
\end{equation}
Here, \(u_0\) is the initial undisturbed-flow velocity, \(t_0\) is the ramp onset time, \(\Delta t_r\) is the ramp duration and \(\Delta u\) is the velocity increment. Unlike an ideal velocity step, the
finite-duration ramp provides a bounded acceleration whose sign, magnitude and duration can be varied systematically. Furthermore, it enables analysis of the history force during both the velocity ramp and the subsequent constant-velocity relaxation. To express the history convolution in terms of the time elapsed since each past acceleration, we introduce the time lag
\begin{equation}
s=t-\tau,
\label{eq:lag_definition}
\end{equation}
where \(\tau\) is the source (past) time at which a past acceleration acted. Thus, \(s=0\) denotes the present time, while increasing \(s\) corresponds to progressively older contributions. The lag is the natural memory coordinate because the kernel weighs each past acceleration according to the time elapsed since it occurred. 

At finite \(\Rey\), we adopt the ansatz that this weighting depends on both the lag and the Reynolds number at the source time \(\tau\), defining
\begin{equation}
a_\infty(t)=\frac{\mathrm{d}u_\infty}{\mathrm{d}t},
\qquad
\Rey(t-s)=\frac{\rho_f\,d\,|u_\infty(t-s)|}{\mu}.
\label{eq:source_time_quantities}
\end{equation}
Under this assumption, the history force can be written as
\begin{equation}
F_{\mathrm H}(t)
=
\Phi\,
\int_{0}^{t}
a_\infty(t-s)\,
K\!\left(s,\Rey(t-s)\right)
\,\mathrm{d}s .
\label{eq:history_force_lag}
\end{equation}
At this stage, \(K(s,\Rey)\) denotes an unknown finite-\(\Rey\) kernel with no functional form imposed. Following Mei (1994), the Reynolds number in \cref{eq:history_force_lag} is evaluated at the source time \(t-s\), rather than at the observation time \(t\). It is important to distinguish the acceleration at the observation time, \(a_\infty(t)\), from the source-time acceleration \(a_\infty(t-s)\) appearing in \cref{eq:history_force_lag}.
For the velocity history in \cref{eq:ramp_velocity}, \(a_\infty(t-s)=\alpha\) only when the corresponding source time lies within the ramp interval:
\begin{equation}
t_0\leq t-s\leq t_0+\Delta t_r.
\end{equation}
Equivalently, the contributing lags satisfy
\begin{equation}
t-(t_0+\Delta t_r)\leq s\leq t-t_0.
\end{equation}
Because the acceleration vanishes outside this interval, the history convolution reduces to a finite-window integral of the kernel over explicitly known lag limits:
\begin{equation}
F_{\mathrm H}(t)
=
\begin{cases}
0,
& t<t_0, \\[2mm]
\displaystyle
\Phi\,\alpha\,
\int_{0}^{t-t_0}
K\!\left(s,\Rey(t-s)\right)
\,\mathrm{d}s,
& t_0\leq t\leq t_0+\Delta t_r, \\[4mm]
\displaystyle
\Phi\,\alpha\,
\int_{t-t_0-\Delta t_r}^{\,t-t_0}
K\!\left(s,\Rey(t-s)\right)
\,\mathrm{d}s,
& t>t_0+\Delta t_r.
\end{cases}
\label{eq:history_force_ramp}
\end{equation}
\Cref{eq:history_force_ramp} identifies two distinct lag-window regimes. During the ramp, the window grows from \(s=0\) as the acceleration history accumulates. After the ramp, the window has the fixed width \(\Delta t_r\) and translates towards progressively larger lags. This distinction motivates the separate in-ramp and post-ramp extraction procedures developed in the next sections.

\subsection{In-ramp kernel extraction} \label{sec:In-ramp extraction}
Let
\begin{equation}
S=t-t_0,
\qquad
0<S<\Delta t_r,
\label{eq:ramp_elapsed_time}
\end{equation}
denote the elapsed time since the onset of the ramp. During this stage, the lower limit of the lag window remains fixed at \(s=0\), while its upper limit increases until \(s=S\). If the kernel depended on lag alone, the history force would be the running cumulative integral
\begin{equation}
F_{\mathrm H}(S)
=
\Phi\,\alpha\,
\int_0^S K(s)\,\mathrm{d}s.
\label{eq:inramp_lag_only}
\end{equation}
Using \(d=2r\), the dimensionless time variables are defined using the viscous timescale as
\begin{equation}
\hat{t}=\frac{\nu_f\,t}{r^2},
\qquad
\hat{S}=\frac{\nu_f\,S}{r^2},
\qquad
\hat{s}=\frac{\nu_f\,s}{r^2},
\qquad
\Delta\hat{t}_r=\frac{\nu_f\,\Delta t_r}{r^2}.
\label{eq:dimensionless_time_variables}
\end{equation}
The same time scaling applies to \(t_0\) and \(\tau\). The corresponding dimensionless acceleration, forcing frequency and history force are
\begin{equation}
\hat{\alpha}=\frac{r^3\,\alpha}{\nu_f^2},
\qquad
\hat{f}=\frac{r^2\,f}{\nu_f},
\qquad
\hat{F}_{\mathrm H}=\frac{F_{\mathrm H}}{6\,\upi\,\mu\,\nu_f}.
\label{eq:dimensionless_validation_variables}
\end{equation}
The velocity scale is \(\nu_f/r\), so that \(\hat{u}=r\,u/\nu_f\) and \(\hat{U}=r\,U/\nu_f\), with the same scaling for velocity increments. Also, \(\hat{a}_\infty=r^3\,a_\infty/\nu_f^2=\mathrm{d}\hat{u}_\infty/\mathrm{d}\hat{t}\) and \(\Rey(\hat{t})=2\,|\hat{u}_\infty(\hat{t})|\). All forces are scaled by \(6\,\upi\,\mu\,\nu_f\). The kernel is dimensionless, so its values are unchanged by this scaling and only its lag argument is rescaled. Hereafter, hats are omitted and all quantities are dimensionless, except for physical parameters and expressions in the appendix. In dimensionless form ~\cref{eq:inramp_lag_only} becomes
\begin{equation}
F_{\mathrm H}(S)
=
\alpha\,
\int_0^{S} K(s)\,\mathrm{d}s, 
\label{eq:inramp_lag_only_dimensionless}
\end{equation}
and the fundamental theorem of calculus gives
\begin{equation}
K(S)
=
\frac{1}{\alpha}\,
\frac{\mathrm{d}F_{\mathrm H}}{\mathrm{d}S}.
\label{eq:inramp_lag_only_extraction}
\end{equation}
This result provides the basic motivation for the derivative-based in-ramp extraction. 

Under the finite-\(\Rey\) ansatz introduced in \cref{eq:history_force_lag}, however, the kernel is evaluated at the Reynolds number corresponding to the source time. The source-time velocity and Reynolds number are
\begin{equation}
u_\infty(t-s)
=
u_0+\alpha\,(S-s),
\qquad
\Rey(t-s)
=
\Rey_0+\Gamma\,(S-s),
\qquad
\Gamma=2\,\alpha,
\label{eq:inramp_source_state}
\end{equation}
where \(\Rey_0\) is the Reynolds number at the ramp onset \(t_0\) and \(\Gamma\) is the rate of change of the Reynolds number during the ramp. The factor of two follows from the diameter-based Reynolds number and the radius-based viscous scales. The in-ramp history force is therefore
\begin{equation}
F_{\mathrm H}(S)
=
\alpha\,
\int_0^{S}
K\!\left(s,\Rey_0+\Gamma\,(S-s)\right)
\,\mathrm{d}s.
\label{eq:inramp_history_force}
\end{equation}
In \cref{eq:inramp_history_force}, \(S\) appears both in the moving upper integration limit and in the Reynolds number argument of the integrand. By applying the Leibniz rule, the same differentiation of the history force now gives
\begin{equation}
\frac{1}{\alpha}\,
\frac{\mathrm{d}F_{\mathrm H}}{\mathrm{d}S}
= {K(S,\Rey_0)} + \Gamma\,\int_0^{S}
\frac{\partial K}{\partial\Rey}
\left(s,\Rey_0+\Gamma\,(S-s)\right)
\,\mathrm{d}s.
\label{eq:inramp_leibniz}
\end{equation}
The first term is the boundary contribution from the moving upper limit. At $s=S$ the source time is $t_0$, so this term is the kernel evaluated at the initial Reynolds number $\Rey_0$. The second term arises from the Reynolds-number dependence of $K$ when the force integral is differentiated. At a fixed lag $s$, the source-time Reynolds number changes as $t$ advances, and the integral accumulates this effect over all lags $0\leq s\leq S$. For ease of notation, we introduce $\mathcal{K}(S;\Rey_0,\alpha) = \frac{1}{\alpha}\,\frac{\mathrm{d}F_{\mathrm H}}{\mathrm{d}S}$.

As we aim to infer $K(s,\Rey_0)$, the first term on the right-hand side of \cref{eq:inramp_leibniz} must be separated from the second term. To do so, we consider ramps of the same duration that start from the same initial Reynolds number \(\Rey_0\),  but have different accelerations. Assuming that the kernel varies smoothly with Reynolds number, \(\partial K/\partial\Rey\) is expanded about \(\Rey_0\):
\begin{equation}
\frac{\partial K}{\partial\Rey}
\left(
s,\Rey_0+\Gamma\,(S-s)
\right)
=
\frac{\partial K}{\partial\Rey}(s,\Rey_0)
+
\Gamma\,(S-s)\,
\frac{\partial^2K}{\partial\Rey^2}(s,\Rey_0)
+
\mathcal{O}\!\left([\Gamma\,(S-s)]^2\right),
\label{eq:kernel_Reynolds_taylor}
\end{equation}
where the expansion is understood at fixed \(S\). Substitution into \cref{eq:inramp_leibniz} gives
\begin{align}
\mathcal{K}(S;\Rey_0,\alpha)
={}&
K(S,\Rey_0)
+
\alpha \,
\underbrace{
2\,
\int_0^{S}
\frac{\partial K}{\partial\Rey}(s,\Rey_0)
\,\mathrm{d}s
}_{D(S,\Rey_0)}
\nonumber\\
&+
\alpha^2 \,
\underbrace{
4\,
\int_0^{S}
(S-s)\,
\frac{\partial^2 K}{\partial\Rey^2}(s,\Rey_0)
\,\mathrm{d}s
}_{Q(S,\Rey_0)}
+
\mathcal{O}(\alpha^3) \,,
\label{eq:inramp_taylor_dimensionless}
\end{align}
which, to reduce notational clutter, we denote by
\begin{equation}
\mathcal{K}(S;\Rey_0,\alpha)
=
K(S,\Rey_0)
+
\alpha\,D(S,\Rey_0)
+
\alpha^2\,Q(S,\Rey_0)
+
\mathcal{O}(\alpha^3)\,,
\label{eq:inramp_taylor_expansion_short}
\end{equation}
with \(\mathcal{O}\) representing the higher-order terms.

\begin{figure}
        \includegraphics[width=1\linewidth]{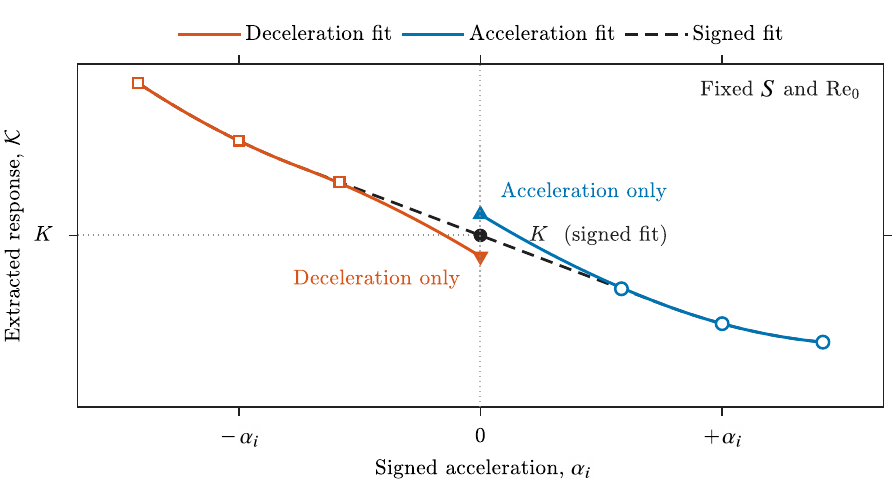}
        \caption{Schematic illustration of signed-ramp extraction. Separate fits to accelerating and decelerating responses yield different estimates at zero acceleration, whereas the joint signed fit gives the intercept \(K\). The curves and symbols do not represent simulation results.}
    \label{fig:paired_epsilon_intercept}
\end{figure}

Next, we parametrise the ramp amplitudes by
\begin{equation}
\epsilon_i=\frac{\Delta u_i}{u_0},
\qquad
\alpha_i=\frac{\Delta u_i}{\Delta t_r}
=\frac{\epsilon_i\,u_0}{\Delta t_r},
\label{eq:epsilon_definition}
\end{equation}
where \(\epsilon_i\) is the dimensionless velocity increment. Positive and negative values correspond to acceleration and deceleration ramps, respectively. For each selected magnitude \(|\epsilon_i|\), matched ramps are performed at identical \(\Rey_0\) and \(\Delta t_r\). Because the accelerations lie on both sides of $\alpha=0$, the zero-acceleration intercept is constrained by acceleration and deceleration ramps rather than by one-sided extrapolation, as shown schematically in \cref{fig:paired_epsilon_intercept}. 

At each fixed lag \(S_j\) and initial Reynolds number \(\Rey_0\), the second-order Taylor expansion motivates a quadratic approximation of the normalised force derivative $\mathcal{K}$ as a function of acceleration. The coefficients $K$, $D$ and $Q$ are determined by least-squares fitting to the numerical results:
\begin{equation}
\begin{bmatrix}
1 & \alpha_1 & \alpha_1^2\\
1 & \alpha_2 & \alpha_2^2\\
\vdots & \vdots & \vdots\\
1 & \alpha_N & \alpha_N^2
\end{bmatrix}
\,
\begin{bmatrix}
K(S_j,\Rey_0)\\
D(S_j,\Rey_0)\\
Q(S_j,\Rey_0)
\end{bmatrix}
\approx
\begin{bmatrix}
\mathcal{K}(S_j;\Rey_0,\alpha_1)\\
\mathcal{K}(S_j;\Rey_0,\alpha_2)\\
\vdots\\
\mathcal{K}(S_j;\Rey_0,\alpha_N)
\end{bmatrix}
\label{eq:inramp_least_squares}
\end{equation}
where the sampled accelerations \(\alpha_i\), \(i=1,\ldots,N\), include both positive and negative values. The fitted constant coefficient, obtained by evaluating the quadratic approximation at \(\alpha=0\), provides an estimate of the in-ramp kernel defined by the theoretical limit
\begin{equation}
\lim_{\alpha\to0}\mathcal{K}(S;\Rey_0,\alpha)
=
K(S,\Rey_0).
\label{eq:zero_acceleration_intercept}
\end{equation}
The velocity-increment dependence shown in \cref{Sensitivity to ramp duration and velocity increment} supports using several ramp amplitudes to determine \(K\). \\

\subsection{Post-ramp kernel extraction} \label{sec:Post-ramp extraction}
After completion of the ramp, \(S=t-t_0>\Delta t_r\), the acceleration at the observation time vanishes. The history force nevertheless remains non-zero because the lag window still contains source times at which the prescribed acceleration was active. From \cref{eq:history_force_ramp} in dimensionless form, the post-ramp history force is
\begin{equation}
F_{\mathrm H}(S)
=
\alpha\,
\int_{S-\Delta t_r}^{S}
K\!\left(
s,\Rey_0+\Gamma\,(S-s)
\right)
\,\mathrm{d}s,
\qquad S>\Delta t_r.
\label{eq:postramp_history_force}
\end{equation}
Unlike the in-ramp extraction, this expression involves a lag window of fixed width \(\Delta t_r\). Defining the midpoint of this window as
\begin{equation}
s_{\mathrm m}
=
S-\frac{\Delta t_r}{2}
=
t-t_0-\frac{\Delta t_r}{2},
\label{eq:postramp_midpoint_lag}
\end{equation}
and using \(\Delta u=\alpha\,\Delta t_r\), the normalised history force becomes
\begin{equation}
\frac{F_{\mathrm H}(S)}{\Delta u}
=
\frac{1}{\Delta t_r}\,
\int_{s_{\mathrm m}-\Delta t_r/2}^{s_{\mathrm m}+\Delta t_r/2}
K\!\left(
s,\Rey_0+\Gamma\,(S-s)
\right)
\,\mathrm{d}s.
\label{eq:postramp_window_average}
\end{equation}
The integral mean-value theorem guarantees that this average equals the kernel evaluated at some lag within the integration interval. However, it does not imply that this lag is the midpoint \((s_{\mathrm m})\). We therefore adopt the midpoint approximation
\begin{equation}
\frac{F_{\mathrm H}(S)}{\Delta u}
\approx
K\!\left(
s_{\mathrm m},
\Rey_0+\frac{\Gamma\,\Delta t_r}{2}
\right).
\label{eq:postramp_midpoint_approximation}
\end{equation}
Unlike the in-ramp derivative, the post-ramp extraction does not generate an explicit accumulated term through Leibniz's rule. The source-time Reynolds number nevertheless varies across the integration window, so the normalised response still depends on the magnitude and sign of \(\alpha\). To obtain $K(s_{\mathrm m},\Rey_0)$, we perform a series of experiments at changing accelerations, and fit the obtained signed post-ramp responses. The sought-after value is obtained as the zero-acceleration fitting:
\begin{equation}
K(s_{\mathrm m},\Rey_0)
\approx
\lim_{\alpha\rightarrow0}
\frac{F_{\mathrm H}(S)}
     {\Delta u}.
\label{eq:postramp_kernel_extraction}
\end{equation}
The midpoint approximation is reliable only when the kernel varies slowly over the averaging window. Accordingly, the post-ramp method is restricted to a lag range in which the kernel response is smooth on the scale of \(\Delta t_r\) and does not exhibit sharp variations. 

\subsection{Combination of in-ramp and post-ramp extraction}
\label{sec:combined_extraction}
To make use of the complementary strengths of the two extraction methods, ramps of two different durations are used. For each starting Reynolds number, \(\Rey_0\), and prescribed velocity change, $\Delta u$, simulations with the same initial and final Reynolds numbers are repeated using ramps of different duration \(\Delta t_r\), with the acceleration adjusted according to \(\alpha=\Delta u/\Delta t_r\).

The longer ramp extends the in-ramp interval \(0<S<\Delta t_r\), allowing the derivative-based method to recover the kernel over both the short and the intermediate lag range. However, this extraction deteriorates at larger lags and for small \(\lvert\epsilon\rvert\), because both the kernel and the acceleration are small, while numerical differentiation amplifies residual noise in the isolated force. Furthermore, using only long ramps would restrict the investigation to small accelerations, because, for a fixed ramp duration \(\Delta t_r\), increasing \(\lvert\alpha\rvert\) leads to larger variations in the Reynolds number, which can easily exceed the range investigated in the present study. 

The post-ramp method avoids differentiation, but returns an average over a lag window of width \(\Delta t_r\). The midpoint value introduces an error, which can be appreciable where the kernel is strongly nonlinear. The short ramp reduces this error by narrowing the averaging window and its post-ramp response is retained only at sufficiently large lags, where the kernel varies slowly across that window. 

The final kernel therefore combines the in-ramp core from the long ramp with the post-ramp tail from the short ramp. This choice is supported by the numerical comparisons in \cref{Sensitivity to ramp duration and velocity increment}, which show only a minor dependence on ramp duration at fixed \(\Rey_0\) and velocity increment within the investigated range. The selected ramp durations and the agreement between the two estimates near the joining point are discussed in \cref{sec:force_evaluation_database_generation}.

\subsection{Treatment of limiting regimes}
\label{sec:Kernel_surface_construction}

The combined in-ramp and post-ramp estimates provide the kernel over a finite range of lags and Reynolds numbers specified in \cref{sec:force_evaluation_database_generation}. Additional treatments are therefore required at short lags, where the extraction becomes unreliable, at Reynolds numbers below the lowest simulated value and at lags beyond the total simulation time. These treatments are formulated in terms of the kernel normalised by the classical Basset kernel,
 \begin{equation}
R(s,\Rey)=\frac{K(s,\Rey)}{K_{\mathrm B}(s)},
\qquad
K_{\mathrm B}(s)=\frac{1}{\sqrt{\upi s}}.
\label{eq:general_normalised_kernel}
\end{equation}
The ratio \(R\) measures the departure from the Basset kernel: \(R=1\) corresponds to the Basset kernel, while \(R<1\) indicates a smaller kernel value at the same lag. Dividing by \(K_{\mathrm B}\) factors out the known \(s^{-1/2}\) singularity. The required short-lag and creeping-flow limits can therefore both be expressed as \(R\rightarrow1\), respectively as \(s\rightarrow0\) at fixed \(\Rey\) and as \(\Rey\rightarrow0\) at fixed \(s\).

\subsubsection{Short-lag treatment}

Direct extraction of \(K(s,\Rey_0)\) becomes increasingly sensitive to errors in force differentiation as \(s\rightarrow0\). The kernel below a prescribed cut-off lag \(s_h\) is therefore replaced by an expression that recovers the classical Basset limit.

We denote the kernel before any closure or constraint is applied by \(K_{\mathrm{raw}}(s,\Rey)\), and define
\begin{equation}
R_{\mathrm{raw}}(s,\Rey)=\frac{K_{\mathrm{raw}}(s,\Rey)}{K_{\mathrm B}(s)},
\qquad
\widetilde R(s,\Rey)=\min\left[R_{\mathrm{raw}}(s,\Rey),1\right].
\label{eq:normalised_kernel_ratio}
\end{equation}
The upper bound represents the adopted constraint that the modified kernel should not exceed \(K_{\mathrm B}\). For \(0<s\leq s_h\), the normalised short-lag kernel is modelled as
\begin{equation}
R_{\mathrm{head}}(s,\Rey)=\exp\left[-A(\Rey)s^q\right],
\label{eq:short_lag_head}
\end{equation}
with
\begin{equation}
A(\Rey)=-\frac{\ln R_h(\Rey)}{s_h^q},
\qquad
R_h(\Rey)=\widetilde R(s_h,\Rey),
\qquad
s_h\simeq8\times10^{-4},
\qquad
q=0.7.
\label{eq:short_lag_head_parameters}
\end{equation}

This functional form satisfies the required short-lag limit. Since \(s^q\rightarrow0\) as \(s\rightarrow0\) for any \(q>0\), the exponent in \cref{eq:short_lag_head} vanishes and \(R_{\mathrm{head}}\rightarrow1\) for every Reynolds number. The definition of \(A(\Rey)\) in \cref{eq:short_lag_head_parameters} also guarantees exact matching to the constrained extracted value at the cut-off, \(R_{\mathrm{head}}(s_h,\Rey)=R_h(\Rey)\).
The cut-off lag is selected based on the observed short-lag noise. The exponent \(q\) controls the curvature with which the normalised kernel approaches unity. A common value is used for all Reynolds numbers, with \(q=0.7\) selected from the short-lag sensitivity analysis. The resulting short-lag kernel is
\begin{equation}
K_{\mathrm{head}}(s,\Rey)=K_{\mathrm B}(s)R_{\mathrm{head}}(s,\Rey).
\label{eq:short_lag_kernel}
\end{equation}

\subsubsection{Low-\texorpdfstring{\(\Rey\)}{Re} treatment}

The numerical database begins at \(\Rey=0.2\), whereas evaluation of the history integral may also require kernel values at \(0\leq\Rey<0.2\). To obtain kernel values in this range, the logarithm of the normalised kernel, \(\ln R\), is interpolated in Reynolds number at each fixed lag \(s\). The interpolation uses the kernel after the short-lag treatment and the upper-bound constraint described above have been applied. In the creeping-flow limit $\Rey\rightarrow0$, the kernel reduces to the classical Basset kernel, which provides the boundary condition
\begin{equation}
R(s,0)=1,
\qquad
\text{equivalently}
\qquad
\ln R(s,0)=0.
\label{eq:low_re_bc}
\end{equation}
Together with the database values of $\ln R$, this condition defines the data for piecewise cubic Hermite interpolation (PCHIP) over $0<\Rey<0.2$. The interpolated values are exponentiated to obtain $R(s,\Rey)$, and the kernel is recovered as
\begin{equation}
K_{\mathrm{low}\textit{Re}}(s,\Rey)=K_{\mathrm B}(s)R(s,\Rey).
\label{eq:low_re_kernel}
\end{equation}
Interpolating $\ln R$ ensures positivity after exponentiation, and the shape-preserving PCHIP formulation avoids the overshoots that can arise with conventional cubic splines.

\subsubsection{Long-lag closure}
\label{sec:long_lag_treatment}

After the short-lag and low-\(\Rey\) treatments, the kernel is available only up to a finite lag \(s_{\mathrm{end}}\), whereas history-force calculations may require memory over much longer times. Setting the kernel to zero beyond this lag would introduce an artificial cut-off. The kernel is therefore continued beyond \(s_{\mathrm{end}}\). We denote the normalised kernel including the preceding treatments by \(R_{\mathrm{surf}}(s,\Rey)\) and define its value at the cut-off as
\begin{equation}
R_{\mathrm{end}}(\Rey)=R_{\mathrm{surf}}(s_{\mathrm{end}},\Rey).
\label{eq:long_lag_handoff_ratio}
\end{equation}
The extended normalised kernel is then
\begin{equation}
R_{\mathrm{ext}}(s,\Rey)=
\begin{cases}
R_{\mathrm{surf}}(s,\Rey), & s\leq s_{\mathrm{end}},
\\[2mm]
R_{\mathrm{cl}}(s,\Rey), & s>s_{\mathrm{end}},
\end{cases}
\label{eq:extended_ratio}
\end{equation}
where \(R_{\mathrm{cl}}\) denotes the long-lag continuation.

For \(\Rey\leq10\), the continuation uses the decay of the Mei–Adrian kernel, expressed as
\begin{equation}
R_{\mathrm{MA}}(s,\Rey)=\frac{K_{\mathrm{MA}}(s,\Rey)}{K_{\mathrm B}(s)}.
\end{equation}
Only its decay shape is retained, while its amplitude is fixed by matching the kernel at the cut-off:
\begin{equation}
R_{\mathrm{LL}}(s,\Rey)=
R_{\mathrm{end}}(\Rey)\,
\frac{R_{\mathrm{MA}}(s,\Rey)}
{R_{\mathrm{MA}}(s_{\mathrm{end}},\Rey)}.
\label{eq:normalised_ma_tail_ratio}
\end{equation}
The subscript \(\mathrm{LL}\) denotes the low-\(\Rey\), long-lag branch. \Cref{eq:normalised_ma_tail_ratio} guarantees \(R_{\mathrm{LL}}(s_{\mathrm{end}},\Rey)=R_{\mathrm{end}}(\Rey)\). At \(\Rey=0\), both \(R_{\mathrm{MA}}\) and \(R_{\mathrm{end}}\) equal unity, so the continuation retains the classical Basset limit.

For \(\Rey\geq50\), the closure is motivated by the far-wake analysis of \citet{Lawrence_Mei_1995}, who specify the far-wake limit of \(s^{-2}\) decay for positive, non-reversing velocity changes but do not specify the finite-lag transition to this regime. We represent this transition using
\begin{equation}
R_{\mathrm{HL}}(s,\Rey)=
R_{\mathrm{end}}(\Rey)\,
\left(\frac{s_{\mathrm{end}}}{s}\right)^{3/2}
\left[
r_\infty+
(1-r_\infty)\exp\left(-\frac{\eta}{L_\eta}\right)
\right],
\qquad
\eta=\frac{u_\infty(s-s_{\mathrm{end}})}{2}.
\label{eq:high_re_tail_ratio}
\end{equation}
The subscript \(\mathrm{HL}\) denotes the high-\(\Rey\), long-lag branch. The variable \(\eta\) measures the convective distance beyond the cut-off in particle diameters, and \(L_\eta\) is the corresponding relaxation length. As \(\eta\rightarrow\infty\), the term in brackets approaches \(r_\infty\), giving \(R_{\mathrm{HL}}\propto s^{-3/2}\) and hence \(K_{\mathrm{HL}}\propto s^{-2}\), because \(K_{\mathrm B}\propto s^{-1/2}\). The complete derivation is given in Appendix \ref{appA}.

The parameter \(r_\infty\) sets the amplitude of the inverse-square tail through
\begin{equation}
r_\infty=
\lim_{s\rightarrow\infty}
\frac{s^{3/2}R(s,\Rey)}
{s_{\mathrm{end}}^{3/2}R_{\mathrm{end}}(\Rey)}.
\label{eq:long_lag_asymptotic_ratio}
\end{equation}
The asymptotic amplitude is estimated from the Lawrence–Mei expression using the numerical steady-drag curve. With \(r_\infty\) fixed, \(L_\eta\) is determined from the rescaled ratio
\begin{equation}
\mathcal{A}=
\frac{R(s,\Rey)}{R_{\mathrm{end}}(\Rey)}
\left(\frac{s}{s_{\mathrm{end}}}\right)^{3/2}.
\label{eq:long_lag_rescaled_ratio}
\end{equation}
Here, \(R\) is obtained from the long-duration ramp calculations. The expression \(r_\infty+(1-r_\infty)\exp(-\eta/L_\eta)\) is fitted to \(\mathcal A\) for each case, and the resulting values of \(L_\eta\) are averaged to obtain a common relaxation length. The adopted parameter values and the comparison with the numerical results are presented in \cref{sec:force_evaluation_database_generation}.

In the intermediate range \(10<\Rey<50\), the two continuations are combined using a logarithmic blend:
\begin{equation}
\ln R_{\mathrm{cl}}=(1-w)\ln R_{\mathrm{LL}}+w\ln R_{\mathrm{HL}},
\qquad
w=\frac{\Rey-10}{40}.
\end{equation}
For \(\Rey\leq10\), \(R_{\mathrm{cl}}=R_{\mathrm{LL}}\), while for \(\Rey\geq50\), \(R_{\mathrm{cl}}=R_{\mathrm{HL}}\). This provides a continuous, positive transition between the two branches. The kernel used in history-force calculations is recovered as \(K(s,\Rey)=K_{\mathrm B}(s)R_{\mathrm{ext}}(s,\Rey)\).

\section{Numerical simulations}
\subsection{Computational setup and numerical method}
\label{sec:computational_setup}

\begin{figure}
    \centering
    \includegraphics[width=1\linewidth]{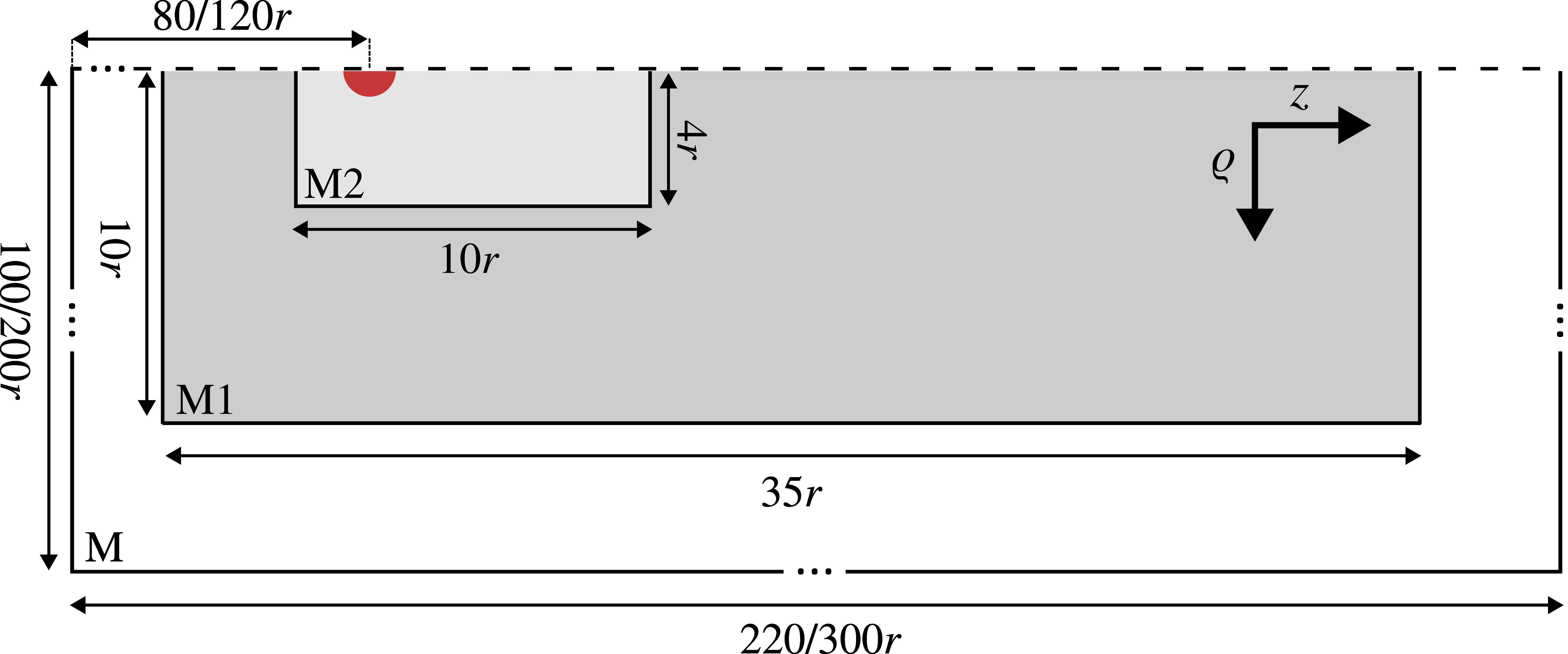}
    \caption{Axisymmetric computational domain and mesh-refinement regions. All dimensions are normalised by the sphere radius \(r\). The first and second values denote the domain dimensions used for \(\Rey\geq1\) and \(\Rey<1\), respectively.}
    \label{fig:domain_schematic}
\end{figure}

Numerical simulations are carried out in COMSOL Multiphysics 6.3 using the finite-element formulation. Because all investigated cases remain below the first symmetry-breaking bifurcation of the sphere wake, which occurs at \(\Rey\approx 212\) \citep{Ern2012WakeInducedOP}, the flow is treated as axisymmetric and swirl-free. In cylindrical coordinates \((\varrho,z)\), the velocity field is expressed as 
\begin{equation}
\boldsymbol{u}
=
u_{\varrho}(\varrho,z,t)\,\boldsymbol{e}_{\varrho}
+
u_z(\varrho,z,t)\,\boldsymbol{e}_z .
\end{equation}
The incompressible Navier-Stokes equations solved are

\begin{subequations}
\label{eq:axisymmetric_NS}
\begin{align}
\frac{1}{\varrho}\,
\frac{\partial (\varrho\,u_{\varrho})}{\partial \varrho}
+
\frac{\partial u_z}{\partial z}
=0,
\label{eq:axisymmetric_continuity}
\\
\frac{\partial\boldsymbol{u}}{\partial t}
+
\boldsymbol{u}\boldsymbol{\cdot}\boldsymbol{\nabla}\boldsymbol{u}
=
\boldsymbol{\nabla}\boldsymbol{\cdot}\boldsymbol{\sigma},
\label{eq:axisymmetric_momentum}
\end{align}
\end{subequations}
where the Cauchy stress tensor is
\begin{equation}
\boldsymbol{\sigma}
=
-p\,\boldsymbol{I}
+
\boldsymbol{\nabla}\boldsymbol{u}
+
\left(\boldsymbol{\nabla}\boldsymbol{u}\right)^{\mathsf T}.
\label{eq:cauchy_stress}
\end{equation}
The computational domain and its dimensions are shown in \cref{fig:domain_schematic}, where \(r\) denotes the sphere radius. The domain is partitioned into regions, \(M\), \(M1\) and \(M2\), with progressively finer resolution towards the sphere. For \(\Rey\geq1\), the dimensions of the outer region \(M\) correspond to the first values reported in \cref{fig:domain_schematic}. The second, larger values are used for \(\Rey<1\). This is because in creeping flow, the leading disturbance velocity decays as \(\mathcal{O}(r/R)\), where \(R\) is the distance from the sphere centre. The influence of the external boundaries therefore decreases only algebraically with their distance from the particle \citep{happel1983low}. The used enlargement accounts for this slow spatial decay of the viscous disturbance near the Stokes limit.  
A spatially uniform velocity
\begin{equation}
\boldsymbol{u} = u_\infty(t)\,\boldsymbol{e}_z
\end{equation}
is prescribed at the inlet, with a constant-pressure condition applied at the outlet. No-slip is enforced on the sphere surface, axisymmetry is imposed at \(\varrho=0\) and slip-wall conditions are applied on the remaining boundary.

The conforming mesh consists predominantly of triangular elements, with quadrilateral boundary-layer elements adjacent to the sphere. Maximum element sizes of \(0.1\,r\) and \(0.05\,r\) are prescribed in \(M1\) and \(M2\), respectively, and \(0.005\,r\) on the sphere surface. Six inflation-layer elements are placed adjacent to the sphere, with a first-layer height of \(5\times10^{-4}\,r\) and a growth factor of \(1.2\). Element size transitions smoothly between regions increasing progressively outside the refinement zones to \(1.6\,r\) for \(\Rey\geq1\) and to \(5.6\,r\) for \(\Rey<1\). The resulting mesh contains approximately \(6.1\times10^4\) elements.

Velocity and pressure are discretised using stabilised equal-order linear finite elements, \(P_1P_1\)  and are solved in a fully coupled manner. Time integration employs a fully implicit, second-order backward differentiation formula. The nonlinear algebraic system is solved using a damped Newton method. The resulting linear systems, comprising approximately \(9.5\times10^4\) degrees of freedom, are solved in double precision using the PARDISO direct solver. 
\subsection{Force evaluation and kernel construction}
\label{sec:force_evaluation_database_generation}
To isolate the history-force contribution, the streamwise hydrodynamic
force is evaluated from the total traction acting on the sphere:
\begin{equation}
F_{\mathrm{num}}
=
\frac{1}{6\upi}
\int_{\mathcal{S}_{\mathrm p}}
\left(
\boldsymbol{\sigma}\boldsymbol{\cdot}\boldsymbol{n}
\right)
\boldsymbol{\cdot}\boldsymbol{e}_z
\,\mathrm{d}A,
\label{eq:streamwise_hydrodynamic_force}
\end{equation}
where \(\mathcal{S}_{\mathrm p}\) denotes the dimensionless particle surface. The prefactor \(1/(6\upi)\) follows from the force normalisation by \(6\upi\mu\nu_f\) defined in \cref{eq:dimensionless_validation_variables}. In the finite-element calculations, \cref{eq:streamwise_hydrodynamic_force} is evaluated using the nodal reaction forces on the sphere boundary. Prior to transient simulations, a stationary parametric sweep is performed over the full range of considered \(u_\infty\). The resulting steady force is tabulated as a function of \(\Rey\), and \(F_{\mathrm{QS}}(\Rey(t))\) is obtained by interpolation. This procedure avoids introducing the uncertainty of a semi-empirical drag correlation into the relatively small unsteady-force residual. Each transient simulation is initialised from the stationary solution corresponding to its initial velocity \(u_0\). Consequently, all terms in the force decomposition \cref{eq:fixed_sphere_force}, except for the history force, are known independently. 

Numerical simulations are performed at 51 initial Reynolds numbers,
spanning \(0.2\leq \Rey_0\leq200\). For each \(\Rey_0\), at least six
signed velocity increments \(\epsilon\) are simulated, comprising at least three
positive and three negative values. Selected cases are also repeated for several ramp durations. Numerical results are subsequently used to extract the kernel up to a dimensionless lag of \(s_{\mathrm{end}}=0.78\), following the procedures described in \cref{sec:In-ramp extraction,sec:Post-ramp extraction}. The isolated history-force signals contain low-level numerical fluctuations, particularly near ramp onset. Since these fluctuations would be amplified by direct differentiation, the derivative
\(\mathrm{d}F_{\mathrm H}/\mathrm{d}S\) is evaluated from weighted
local quadratic fits in logarithmic time using tricube weights.

\begin{figure}
    \centering
    \includegraphics[width=0.95\linewidth]{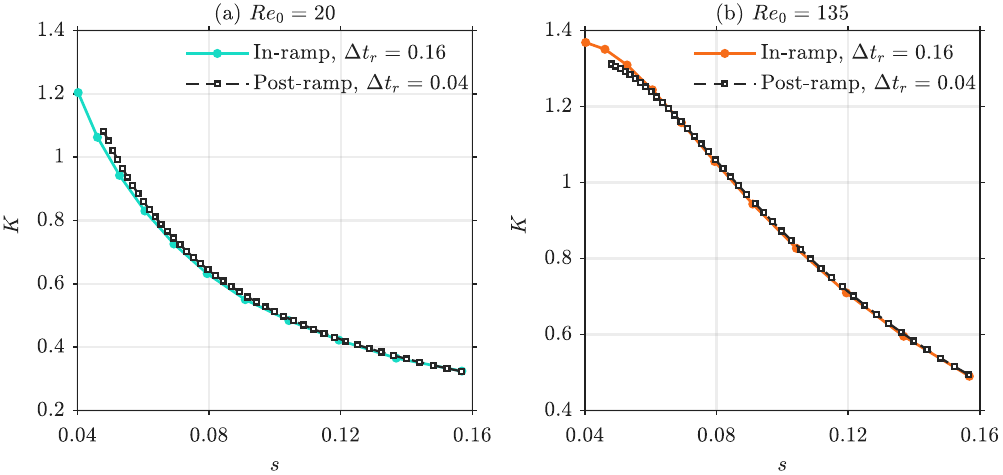}
    \caption{Overlap between the long-ramp in-ramp core and the short-ramp post-ramp tail. Left: \(\Rey_0=20\). Right: \(\Rey_0=135\).}
    \label{fig:core_tail_overlap}
\end{figure}

The in-ramp procedure is applied to the long ramp of duration \(\Delta t_r=0.16\), whereas the post-ramp procedure is applied to the short ramp of duration \(\Delta t_r=0.04\). The two contributions are combined following \cref{sec:combined_extraction}. The two methods agree closely in the overlap region, providing an internal consistency check, as illustrated in \cref{fig:core_tail_overlap}.

Additional long-duration acceleration and deceleration simulations are performed for several Reynolds numbers and velocity increments to support the long-lag closure described in \cref{sec:long_lag_treatment}. The kernels extracted from these calculations are used to determine the closure parameters and assess the resulting continuation, but are not added to \(R_{\mathrm{surf}}\).

For \(\Rey\geq50\), the long-duration simulations resolve a rapidly decaying pre-asymptotic response, whereas the ultimate inverse-square regime cannot be resolved directly because the kernel becomes comparable to the numerical noise at large lags. The asymptotic decay is therefore supplied by the Lawrence–Mei analysis, as described in \cref{sec:long_lag_treatment}.

\begin{figure}
    \centering
    \includegraphics[width=\linewidth]{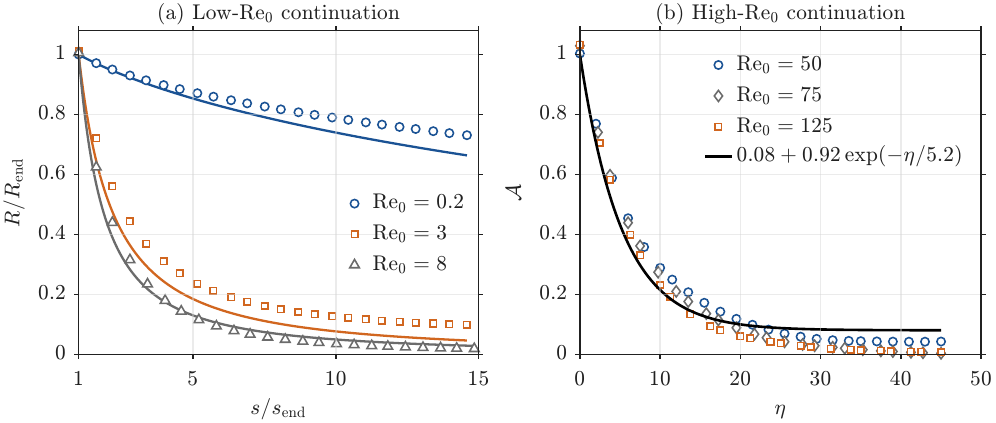}
    \caption{Comparison of the long-lag continuations with kernels extracted from long-duration numerical simulations. Symbols denote the extracted kernels. (a) Low-\(\Rey_0\) data shown as \(R/R_{\mathrm{end}}\) against \(s/s_{\mathrm{end}}\), with solid curves showing the normalised Mei–Adrian continuation. (b) Rescaled high-\(\Rey_0\) data, \(\mathcal{A}=(R/R_{\mathrm{end}})\,(s/s_{\mathrm{end}})^{3/2}\), plotted against the wake-convected distance \(\eta=u_\infty\,(s-s_{\mathrm{end}})/2\). The solid curve shows the adopted relaxation with \(r_\infty=0.08\) and \(L_\eta=5.2\).}
    \label{fig:Tail_Closure_Regime_Validation}
\end{figure}

The asymptotic amplitude ratios are close to \(0.08\) across the examined high-\(\Rey\) cases, motivating the common value \(r_\infty=0.08\). With \(r_\infty\) fixed, \(L_\eta\) is fitted for each case as described in \cref{sec:long_lag_treatment}, giving an average value of \(L_\eta=5.2\). \Cref{fig:Tail_Closure_Regime_Validation} compares the adopted continuations with the kernels extracted from the long-duration calculations. The largest relative deviations occur at the smallest kernel values.

\subsection{Sensitivity to ramp duration and velocity increment} 
\label{Sensitivity to ramp duration and velocity increment}
\begin{figure}
    \centering
    \includegraphics[width=1\linewidth]{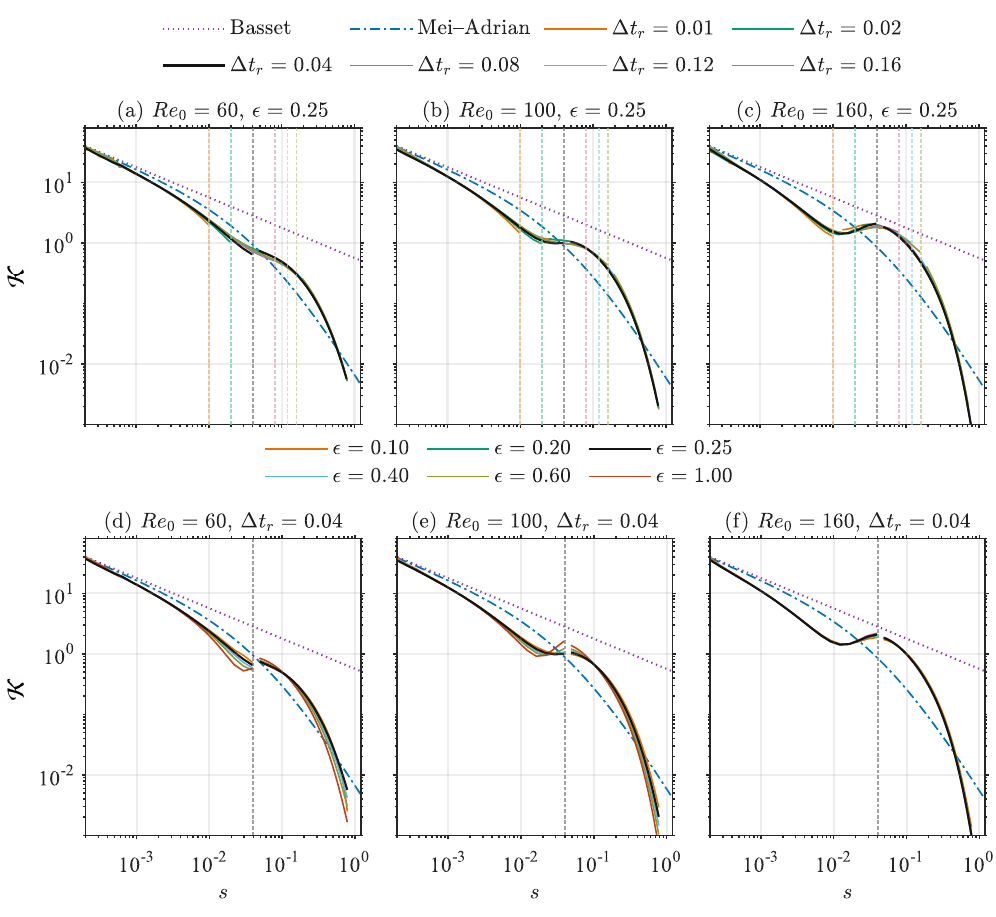}
    \caption{Normalised in-ramp force derivatives and post-ramp kernel estimates illustrating the effects of ramp duration and velocity increment. Panels (a)–(c) show estimates with fixed \(\epsilon\) and varying \(\Rey_0\). Panels (d)–(f) show varying positive \(\epsilon\) at fixed \(\Delta t_r=0.04\). The vertical lines mark the corresponding ramp  ends. Post-ramp estimates are shown only for \(s\geq1.2\,\Delta t_r\). The interval around ramp termination is excluded because neither the in-ramp derivative nor the early post-ramp window estimate is sufficiently reliable there. The classical Basset kernel and the Mei–Adrian kernel evaluated at \(\Rey_0\) are included as reference curves.} 
    \label{fig:velocity_sensitivity}
\end{figure}

We examine how the response obtained from a finite ramp depends on the parameters of the imposed velocity change. \Cref{fig:velocity_sensitivity} compares two complementary parameter sweeps at \(\Rey_0= 60\), 100 and 160. The upper row varies the ramp duration at fixed \(\epsilon=0.25\), whereas the lower row varies the velocity increment at fixed \(\Delta t_r=0.04\). The case \(\Delta t_r=0.04,\epsilon=0.25\) is common to both rows and provides a direct reference between the two sweeps. In the figure, \(\mathcal{K}\) denotes the normalised history-force derivative during the ramp and the post-ramp kernel estimate based on the midpoint approximation, given by \cref{eq:inramp_leibniz,eq:postramp_midpoint_approximation}, respectively. These quantities are obtained from individual ramps before the fitting procedure used to determine \(K\).

At sufficiently short lags, all responses approach the classical Basset kernel, consistent with the dominance of viscous diffusion. The departure from this limit occurs at progressively smaller dimensionless lag as \(\Rey_0\) increases. At intermediate lag, the initial decay develops into an increasingly pronounced shoulder or hump near \(s\approx0.04\). Its position remains approximately unchanged as \(\Delta t_r\) is varied, while the ramp-end markers move across the lag axis. This suggests that the feature is not solely caused by acceleration switch-off, but might be associated with a wake-related response. At larger lags, the curves lie below the classical Basset and Mei-Adrian kernels. 

The upper row examines the effect of ramp duration. At fixed \(\epsilon\), changing \(\Delta t_r\) changes the dimensionless acceleration according to \(\alpha=\epsilon\,\Rey_0/(2\,\Delta t_r)\) while preserving the initial and final Reynolds numbers. The close agreement between the curves indicates little sensitivity to ramp duration within the investigated range. This observation should not be interpreted as general independence of acceleration because the acceleration magnitude and its duration are varied together.

A different behaviour is obtained when \(\epsilon\) is varied at fixed \(\Delta t_r\). In this case, the acceleration and velocity increment change together, altering both the final Reynolds number and the source-time Reynolds-number history. The results diverge in the intermediate and large lag regions.  At \(\Rey_0=60\) and \(100\), increasing \(\epsilon\) generally reduces the response on the pre-hump branch, whereas this ordering weakens and may reverse near the hump. At \(\Rey_0=160\), the sweep is restricted to \(\epsilon=0.1\), \(0.2\) and \(0.25\) to remain within the investigated Reynolds number range. The smaller separation in this panel should therefore be interpreted with caution.

These comparisons support two aspects of the extraction methodology. The limited sensitivity to ramp duration supports using different durations for the in-ramp and post-ramp procedures. The systematic dependence on velocity increment supports determining \(K\) from several ramp amplitudes, following the fitting procedures described in \cref{sec:In-ramp extraction,sec:Post-ramp extraction}.

\section{Kernel structure and wake dynamics}
\subsection{Reynolds-number-dependent memory regimes}
\begin{figure}
    \centering
    \includegraphics[width=1\linewidth]{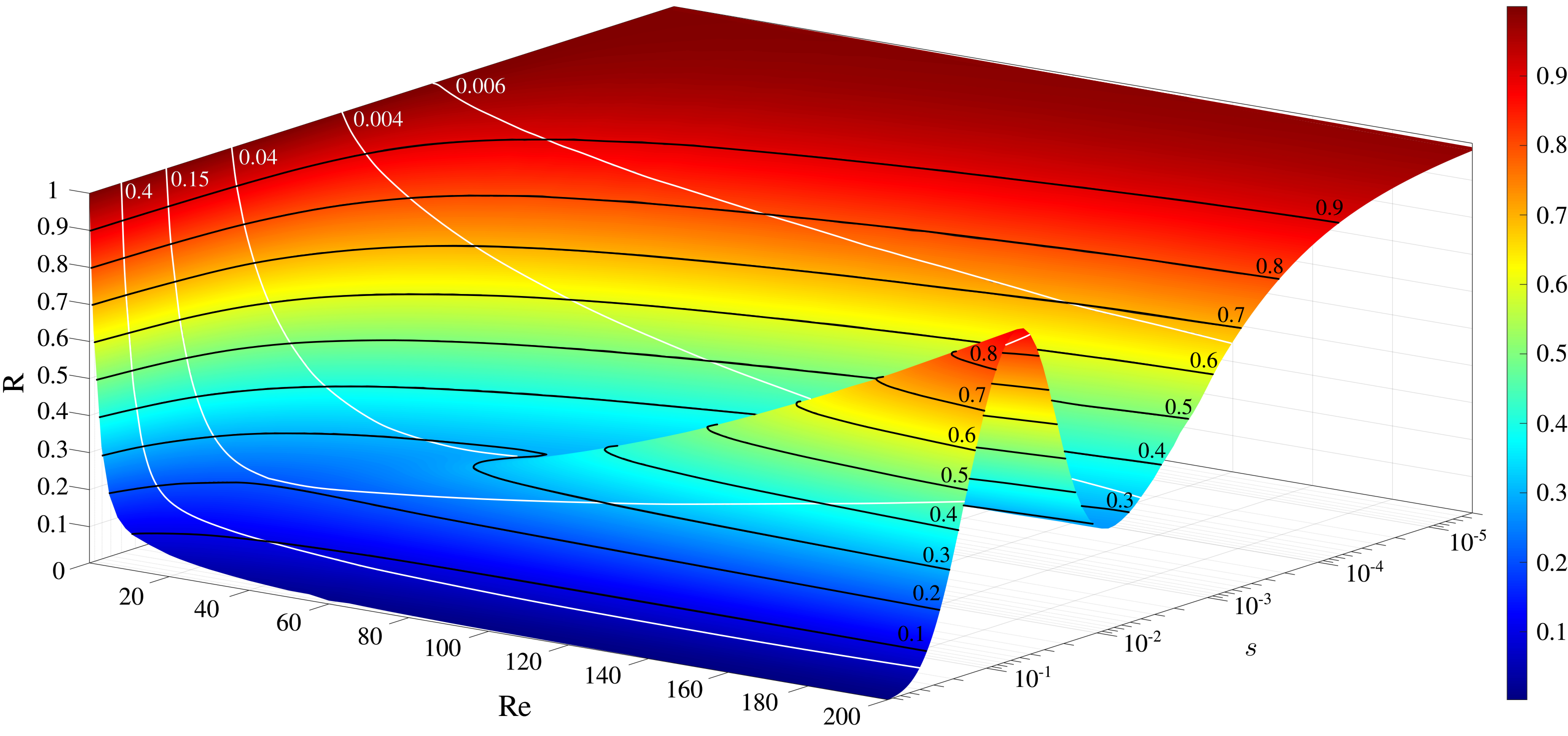}
    \caption{Reconstructed kernel surface normalised by the classical
    Basset kernel, \(R=K/K_{\mathrm B}\), as a function of Reynolds number and
    dimensionless lag \(s\). Black curves are isolines of \(R\),
    and white curves mark selected values of \(s\).}
    \label{fig:kernel_surface}
\end{figure}

Figure~\ref{fig:kernel_surface} presents the completed normalised kernel surface \(R=K/K_{\mathrm B}\). Thus, \(R=1\) corresponds to the Basset kernel, whereas \(R<1\) represents a reduction of the memory kernel at finite Reynolds number. The black curves denote isolines of \(R\), while the white curves indicate selected values of the dimensionless lag \(s\). 

At the shortest displayed lags, the surface forms an approximately Reynolds number independent ridge at \(R\simeq1\), corresponding to the common Basset short-lag limit. A small departure from unity develops for increasing \(\Rey\), with \(R\) decreasing to approximately \(0.988\) at the highest Reynolds number. This trend indicates that convection begins to influence the response at progressively shorter lags as \(\Rey\) increases, so that the purely diffusive regime at the highest \(\Rey\) is confined to times below the displayed range, approximately \(s < 4 \times10^{-6}\). This behaviour is qualitatively consistent with the short-lag scaling of the MA kernel. The precise magnitude of the departure from unity, however, depends on the imposed short-lag treatment and the selected value of \(q\).

As the lag increases, the kernel increasingly deviates from the Basset limit, and the observed reduction becomes more strongly dependent on the Reynolds number. At intermediate lags, approximately between \(s=0.004\) and \(0.15\), the initially smooth decay changes shape. A weak shoulder at lower Reynolds number develops into a pronounced hump as \(\Rey\) increases. The hump-like event is also reported by \citet{Mei_Lawrence_Adrian_1991} and \citet{WAKABA_2005}, but not analysed in depth. The change in the surface shape indicates that, for higher \(\Rey\) the kernel cannot be represented by the MA type formulation and requires a different functional form. Beyond the hump, the surface declines at a Reynolds-number-dependent rate. At the largest dimensionless lag considered here, \(s=0.78\), the normalised kernel reaches approximately \(R=0.06\) , \(0.009\), and \(0.002\) for \(\Rey=20\), \(80\), and \(200\) respectively.

\subsection{Wake dynamics associated with kernel non-monotonicity}

The change from a monotonically decaying kernel at low Reynolds number to an intermediate lag shoulder or hump at higher Reynolds number suggests a corresponding change in the transient wake dynamics. We examine this connection in three steps: first through matched-steady pressure and vorticity fields, then through the regional vorticity balance and finally through the pressure and viscous contributions to the force. 

The kernel-surface analysis indicates that the normalised kernel begins to depart from the Mei-Adrian shape at \(\Rey\approx40\), followed by a transition regime and a developed hump in \(R\) at \(\Rey\approx120\). To isolate the transient flow response, the steady solution at the corresponding instantaneous free-stream velocity is subtracted from each transient solution. The flow-field analysis uses acceleration and deceleration ramps with \(\epsilon=\pm0.25\) and \(\Delta t_r=0.16\). The normalised vorticity difference is therefore defined as
\begin{equation}
    \Delta\omega_{\phi}^{*}(\boldsymbol{x},t)
    =
    \left[
        \omega_\phi(\boldsymbol{x},t)
        -
        \omega_{\phi,\mathrm{ss}}
        \bigl(\boldsymbol{x},u_\infty(t)\bigr)
    \right]\,
\frac{2}{u_\infty(t)},
\label{eq:vorticity_difference}
\end{equation}
where the already dimensionless vorticity difference is divided by \(u_\infty(t)/2\) to facilitate comparison across Reynolds numbers, with 2 representing the dimensionless particle diameter.
The corresponding normalised pressure difference is
\begin{equation}
    \Delta p^{*}(\boldsymbol{x},t)
    =
    \frac{p(\boldsymbol{x},t)
    -p_{\mathrm{ss}}\bigl(\boldsymbol{x},u_\infty(t)\bigr)}
    {2\lvert \alpha \rvert}.
    \label{eq:pressure_difference}
\end{equation}
Here, the subscript \(\mathrm{ss}\) denotes the steady solution at the instantaneous velocity \(u_\infty(t)\). The pressure difference is additionally normalised by \(2\lvert\alpha\rvert\), the dimensionless pressure scale associated with the acceleration imposed during the ramp. Normalisation by the dynamic pressure \(u_\infty^2/2\) would make the normalised values depend strongly on velocity, resulting in substantially different ranges across the Reynolds-number cases. The acceleration-based normalisation reduces this variation and therefore allows the transient pressure structures to be compared over similar numerical ranges. The distance travelled since ramp onset is defined as
\begin{equation}
\chi(t)
=
\int_{t_0}^{t} u_\infty(\tau)\,\mathrm{d}\tau.
\label{eq:hump_convected_distance}
\end{equation}
The distance \(\chi(t)/2\), measured in particle diameters, provides a dimensionless metric of convective time. Interestingly, for \(\Rey=80,100,120,140\) and \(160\), the shoulders and humps are centred at \(\chi(t)/2\approx1\) for acceleration and \(\chi(t)/2\approx0.95\) for deceleration. For each Reynolds number, the shoulder or hump centre is defined as the minimum of the local logarithmic decay exponent, \(-\mathrm{d}\log K/\mathrm{d}\log s\). 

\begin{figure}
    \centering
    \includegraphics[width=1\linewidth]{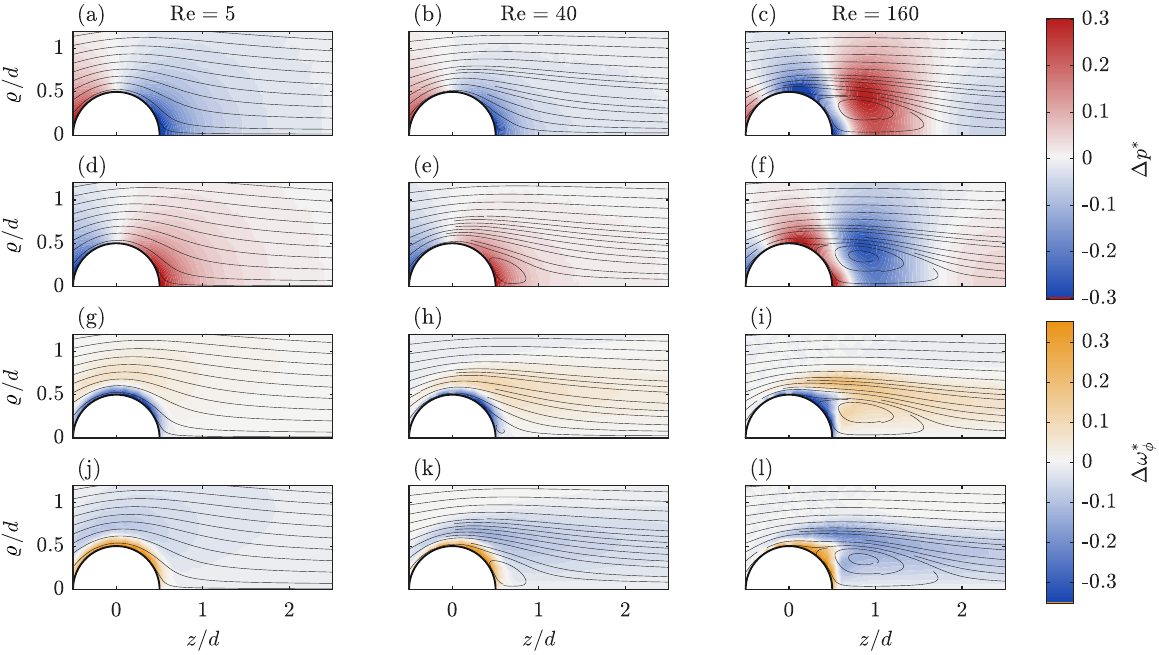}
     \caption{Matched-steady pressure and vorticity differences at \(S=0.026\) for \(\Rey=5\), 40 and 160. Panels (a-f) show \(\Delta p^{*}\), and panels (g-l) show \(\Delta\omega_{\phi}^{*}\), with acceleration above deceleration. Thin black lines denote streamlines of the total velocity field.} 
    \label{fig:jfm_hump_vorticity}
\end{figure}

\Cref{fig:jfm_hump_vorticity} shows the vorticity and pressure differences at three Reynolds numbers for both acceleration and deceleration. The fields are shown at \(S=0.026\), corresponding to the centre of the developed-hump interval at \(\Rey=160\).

At \(\Rey=5\), the vorticity difference is strongest near the sphere surface, where it is distributed relatively uniformly, and has little downstream reach. The pressure difference forms a broad near-body pattern, with regions on both the upstream and downstream sides of the sphere.
At \(\Rey=40\), the vorticity difference becomes more concentrated on the downstream side of the sphere, indicating a more pronounced wake-related response. The pressure difference still maintains a structure similar to that observed at lower Reynolds number.
At \(\Rey=160\), the vorticity difference follows the separated shear layer and develops into a near-wake structure. This change is accompanied by a wake-centred pressure difference behind the sphere. The recordings show that this pressure structure persists in the near wake for a long period after the velocity ramp has ended. These observations suggest a reorganisation of the separated shear layer, near-wake vorticity and pressure distribution around the rear of the sphere.

The acceleration and deceleration fields have approximately opposite signs and similar spatial distributions, although some differences in the shape and position of the near-wake structures become visible at \(\Rey=160\). A separate velocity-field analysis, using the same initial steady solution \(\boldsymbol{u}_0\) for both ramp directions, confirms that the velocity perturbations remain nearly antisymmetric around the shoulder or hump interval. The normalised symmetry defect is very small during this interval but increases with \(\Rey\) and time. 

To examine the development of these structures more deeply we consider an azimuthal vorticity balance. For axisymmetric flow without swirl, the vorticity equation is
\begin{equation}
\underbrace{\frac{\partial\omega_\phi}{\partial t}}_{T_t}
=
\underbrace{
-u_{\varrho}\,\frac{\partial\omega_\phi}{\partial\varrho}
-u_z\,\frac{\partial\omega_\phi}{\partial z}
}_{T_A}
+
\underbrace{\frac{u_{\varrho}\,\omega_\phi}{\varrho}}_{T_S}
+
\underbrace{
\,
\frac{\partial^2\omega_\phi}{\partial\varrho^2}
+\frac{1}{\varrho}\,\frac{\partial\omega_\phi}{\partial\varrho}
+\frac{\partial^2\omega_\phi}{\partial z^2}
-\frac{\omega_\phi}{\varrho^2}
}_{T_D},
\label{eq:vorticity_transport_hump}
\end{equation}
Here \(T_t\) \(T_A\), \(T_S\) and \(T_D\) denote accumulation, advection, axisymmetric stretching and diffusion. In axisymmetric flow, the vortex-stretching term reduces to \(u_{\varrho}\,\omega_\phi/\varrho\) and describes how the radial fluid motion strengthens or weakens vorticity. The transport terms are integrated over fixed geometric regions. The wake core is \(\mathcal{W}_c=\{1<z\leq4\,,\ 0\leq\varrho\leq0.8\}\). This region encloses the central separated wake at \(\Rey=160\), but its name does not imply that recirculation exists at \(\Rey=20\). The wider near-wake region \(\mathcal{W}_n=\{1<z\leq4,\ 0\leq\varrho\leq 2\}\) contains the core and the surrounding shear layer. In both definitions, the first coordinate range specifies the axial extent along \(z\), measured downstream from the sphere centre, whereas the second specifies the radial extent \(\varrho\), measured from the symmetry axis.

\begin{figure}
    \centering
    \includegraphics[width=\linewidth]{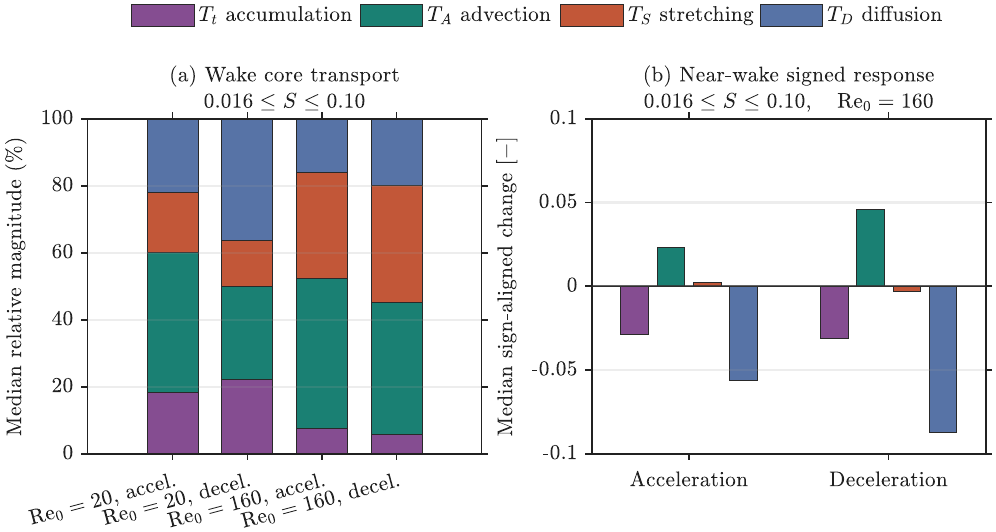}
    \caption{Vorticity-budget diagnostics over \(0.016 \leq S \leq 0.10\). (a) Median relative magnitudes of the accumulation, advection, stretching and diffusion terms in the wake core for acceleration and deceleration at \(\Rey_0=20\) and \(160\). (b) Median near-wake changes at \(\Rey_0=160\), relative to the pre-ramp values and multiplied by the sign of the imposed acceleration.}
    \label{fig:vorticity_balance}
\end{figure}

\Cref{fig:vorticity_balance} summarises the vorticity budget results using two complementary diagnostics. Panel~(a) reports the median relative magnitudes of the accumulation, advection, stretching and diffusion terms in the wake core over \(0.016 \leq S \leq 0.10\). At each time, the volume integral of the absolute value of each term is divided by the sum of these integrals for all four terms. The median of each resulting share is then taken over the stated interval. Because this measure is constructed from the absolute magnitudes of the individual terms, it identifies their relative importance but does not retain their signs or describe  cancellation between them. Averaging the acceleration and deceleration cases at each Reynolds number, the stretching contribution increases from \(15.8\,\%\) at \(\Rey_0=20\) to \(33.4\,\%\) at \(\Rey_0=160\), whereas the diffusion contribution decreases from \(29.1\,\%\) to \(18.0\,\%\). Moreover, the accumulation contribution decreases from approximately \(20\,\%\) to \(7\,\%\), while advection remains one of the dominant terms. Advection and stretching therefore account for approximately three quarters of the wake-core relative magnitude at \(\Rey_0=160\), compared with approximately one half at \(\Rey_0=20\). The hump cases are thus associated with a redistribution of the wake-core budget from accumulation and diffusion towards advection and stretching. At \(\Rey_0=20\), reversing the ramp direction primarily redistributes the relative magnitude between advection and diffusion. At \(\Rey_0=160\), the predominance of advection and stretching is retained for both ramp directions.

Panel~(b) reports the median changes in the near-wake budget terms relative to their pre-ramp values over the same time interval. To compare the two directions on a common basis, the change is multiplied by the sign of the imposed acceleration. The median advective change is positive for both ramp directions, whereas the median accumulation and diffusive changes are negative. The stretching change is comparatively small and changes sign between acceleration and deceleration. The magnitudes of the advective and diffusive changes are larger during deceleration, with diffusion exhibiting the largest median change. Acceleration and deceleration therefore retain the same dominant advection-diffusion sign structure, although the near-wake changes are larger during deceleration.

The flow and transport fields suggest a delayed wake adjustment, but they do not by themselves establish how this adjustment enters the hydrodynamic force. To make this connection, the kernel is decomposed into pressure and viscous contributions, \(K(s)=K_p(s)+K_\mu(s)\). The pressure and viscous history forces are obtained from the total pressure and viscous forces by subtracting the corresponding matched-steady forces. Because the pressure-gradient and added-mass forces also act through surface pressure, they are subtracted from the pressure force as well, consistent with the added-mass analysis of \citet{Chang_Maxey_1995}. The extraction methodology described in \cref{sec:In-ramp extraction} is applied separately to the pressure and viscous history-force signals, giving \(K_p\) and \(K_\mu\) as the zero-acceleration intercepts.

At \(\Rey_0=20\), both the pressure and viscous contributions decrease throughout the displayed lag interval, and the kernel therefore decays monotonically. At \(\Rey_0=160\), the viscous contribution continues to decrease, whereas the pressure contribution rises from an initially negative value and becomes positive. Over the rising part of the hump, the rate of increase of the pressure contribution temporarily exceeds the rate of decrease of the viscous contribution, \(\mathrm{d}K_p/\mathrm{d}s > -\mathrm{d}K_{\mu}/\mathrm{d}s\), so that the kernel increases. The kernel reaches its maximum when the two rates balance and resumes its decay once the pressure increase can no longer compensate for the continuing viscous decrease.

Taken together, the results support the following interpretation. The imposed velocity change initially creates a departure of the transient boundary layer and wake from the corresponding steady flow at the same instantaneous Reynolds number. At \(\Rey_0=20\), the response is associated with larger relative contributions from accumulation and diffusion, and neither \(K_p\) nor \(K_\mu\) develops an increase capable of interrupting the decay of \(K\). At \(\Rey_0=160\), advection and axisymmetric stretching account for a larger fraction of the wake-core term magnitudes, while the near-wake response is characterised by opposing advective and diffusive changes. This delayed wake reorganisation is consistent with the increase in the pressure contribution \(K_p\). When its rate of increase exceeds the rate of decrease of \(K_\mu\), the kernel develops the intermediate-lag shoulder or hump.
\begin{figure}
    \centering
    \includegraphics[width=0.95\linewidth]{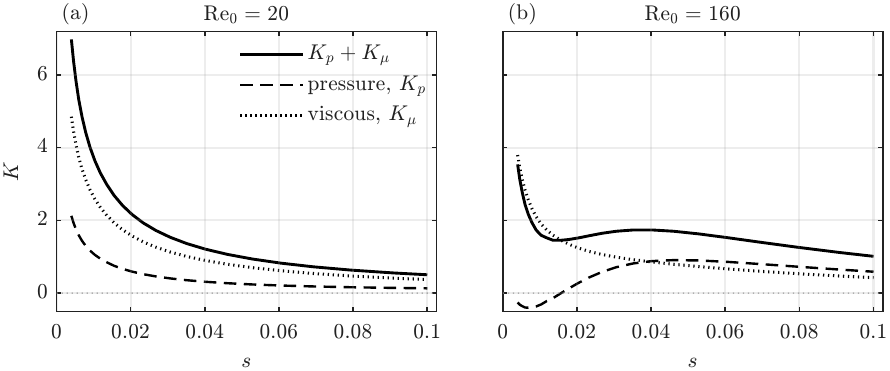}
    \caption{Pressure and viscous contributions to the kernel \(K\) at (a) \(\Rey_0=20\) and (b) \(\Rey_0=160\). Dashed and dotted lines show \(K_p\) and \(K_\mu\), respectively, while the solid line denotes their sum.}
    \label{fig:kernel_force_components}
\end{figure}

\section{Kernel validation}
In this section, we assess whether the surface can reconstruct history-force signals for additional velocity histories without case-specific adjustment. In every comparison, the present surface and the finite-\(\Rey\) literature kernels are evaluated at the Reynolds number corresponding to the source time, while all formulations are convolved with the same prescribed acceleration history. To handle the short-lag singularity, the change of variable \(s=\zeta^2\), with \(\mathrm{d}s=2\,\zeta\,\mathrm{d}\zeta\), is applied before numerical quadrature. Since \(K(s,\Rey)\sim1/\sqrt{\upi s}\) as \(s\to0\), the transformed factor \(2\,\zeta\,K(\zeta^2,\Rey)\) remains finite as \(\zeta\to0\).

\subsection{Finite-ramp histories}
\begin{figure}
    \centering
    \includegraphics[width=1\linewidth]{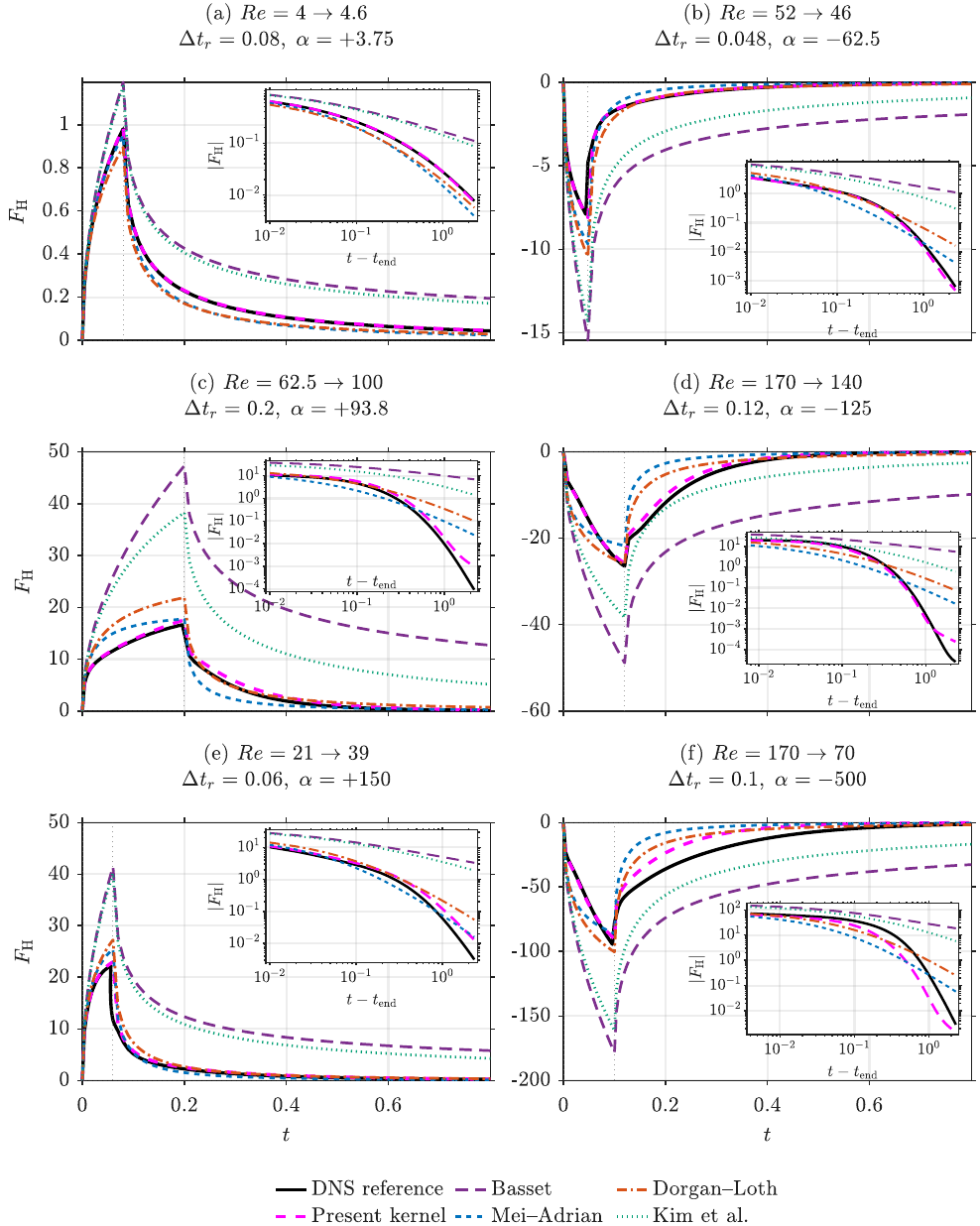}
    \caption{Dimensionless history-force reconstruction \(F_{\mathrm H}(t)\) for six finite ramps covering acceleration and deceleration in different flow conditions, with \(t=0\) at ramp onset. The vertical dotted line marks the end of the ramp, and the insets show the magnitude \(\lvert F_{\mathrm H}\rvert\) of the post-ramp relaxation against \(t-t_{\mathrm{end}}\), where \(t_{\mathrm{end}}\) is the dimensionless ramp-end time, on logarithmic axes.}
    \label{fig:ramp_validation}
\end{figure}

Finite ramps provide the closest validation to the developed extraction method, while allowing the initial Reynolds number, ramp direction, duration, and velocity increment to differ from those used to construct the surface. They therefore test whether the present kernel can be interpolated in Reynolds number and applied to finite velocity changes without case-specific adjustment.

\Cref{fig:ramp_validation} compares six acceleration and deceleration ramps spanning \(4 \leq \Rey \leq 170\), \(0.048 \leq \Delta t_r \leq 0.2\), and \(3.75 \leq |\alpha| \leq 500\). Among the formulations compared, the present surface gives the closest overall agreement with the numerical results across all six cases, capturing the sign, timing and magnitude of the force extremum, and the subsequent relaxation. Its largest discrepancies occur during the post-ramp decay, particularly in cases with the largest acceleration \(|\alpha|\). The classical Basset kernel generally overpredicts the force magnitude and retains excessive long-time memory, whereas the finite-\(\Rey\) literature kernels do not consistently reproduce both the force extremum and the subsequent decay.

At the latest times in panels~(c) and~(f), the surface prediction develops a slight upward curvature in the logarithmic tail. For these cases, the source-time Reynolds numbers remain above \(\Rey=50\), so the long-lag contribution is evaluated using the high-\(\Rey\) closure described in \cref{sec:long_lag_treatment}. In this closure, the common value \(r_\infty\) fixes the ultimate inverse-square amplitude relative to its value at the cut-off, while \(L_\eta\) specifies the corresponding convective relaxation distance. Since these parameters are applied globally, the detailed shape of the final tail remains sensitive to the prescribed continuation. However, the history-force magnitude is already small relative to its ramp-associated extremum, so the related discrepancy has only a minor effect in absolute terms.

\subsection{Oscillatory histories}

Sinusoidal velocity histories provide a further test of the present kernel. Unlike a single ramp, they repeatedly revisit the same Reynolds-number range during alternating acceleration and deceleration phases, causing memory contributions from successive half-cycles to overlap. 

\begin{figure}
    \centering
    \includegraphics[width=1\linewidth]{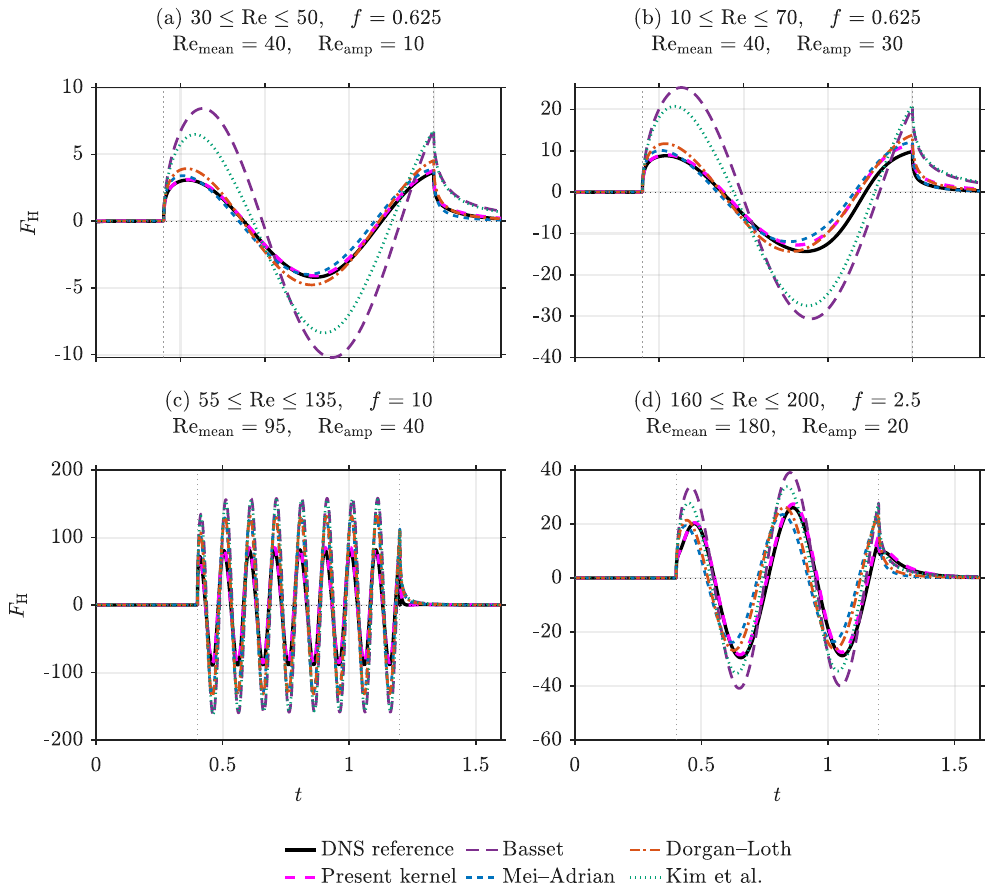}
    \caption{Dimensionless history-force reconstruction \(F_{\mathrm H}(t)\) for four different sinusoidal velocity histories. The vertical dotted lines mark the beginning and end of the oscillatory interval.}
    \label{fig:kernel_surface_oscillating}
\end{figure}

\Cref{fig:kernel_surface_oscillating} compares four cases covering \(10 \leq \Rey \leq 200\) and \(0.625\leq f\leq10\). Here, \(\Rey_{\mathrm{mean}}\) and \(\Rey_{\mathrm{amp}}\) denote the mean and amplitude of the Reynolds-number variation, respectively. In panels~(a), (c), and~(d), the present kernel surface closely reproduces the amplitude and phase of the numerical history force throughout the oscillatory interval, with the predicted and numerical curves nearly indistinguishable at the scale of the figure. The subsequent relaxation after the oscillation ends is also accurately captured.

A larger discrepancy is observed in panel~(b), which combines the lowest frequency with the largest relative oscillation amplitude. The prediction from the present kernel initially follows the numerical response but begins to diverge as the negative force minimum is approached. The resulting errors in both amplitude and phase persist through the remainder of the oscillatory interval. The comparison between panels~(a) and~(b) is particularly informative because both cases have \(\Rey_{\mathrm{mean}}=40\) and \(f=0.625\), while \(\Rey_{\mathrm{amp}}\) increases from \(10\) to \(30\). This increase broadens the sampled Reynolds-number interval from \(30\leq\Rey\leq50\) to \(10\leq\Rey\leq70\) and increases the acceleration amplitude by a factor of three. This comparison indicates that the low forcing frequency alone cannot explain the reduced accuracy in panel~(b).
\begin{figure}
    \centering
    \includegraphics[width=1\linewidth]{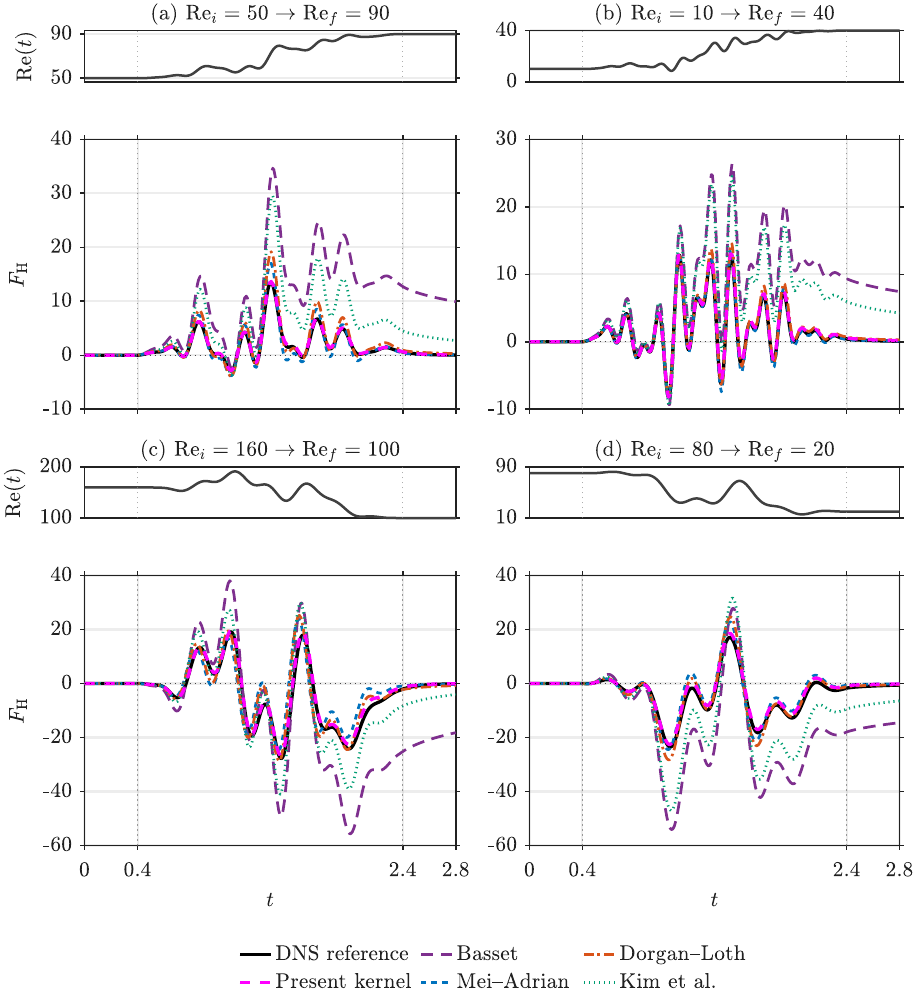}
    \caption{Dimensionless history-force reconstruction \(F_{\mathrm H}(t)\) for four non-periodic composite velocity histories.
    Panels (a) and (b) have a net increase in Reynolds number, whereas
    panels (c) and (d) have a net decrease. The upper subpanels show \(\Rey(t)\), with panel headings giving the initial and final values \(\Rey_i\) and \(\Rey_f\), and the lower subpanels compare the numerical history force with the model predictions.}    \label{fig:kernel_surface_four_case_semirealistic_validation}
\end{figure}
The literature kernels exhibit larger errors in the force amplitude, phase and residual memory. Nevertheless, the Mei-Adrian and Dorgan-Loth formulations remain reasonably accurate and perform more favourably here than in the finite-ramp validation. 

\subsection{Composite histories}

The last validation case uses a more representative behaviour for particle-laden flow than the two fundamental cases investigated before. Each velocity history is constructed by combining a finite ramp with several sinusoidal components of different amplitudes, frequencies, and phases.

\Cref{fig:kernel_surface_four_case_semirealistic_validation} presents two histories with a net increase in Reynolds number, panels~(a) and~(b), and two with a net decrease, panels~(c) and~(d). Among the formulations considered, the present kernel provides the closest overall agreement with the numerical results. It closely reproduces the timing and magnitude of the successive positive maxima and negative minima, including the densely spaced sequence of extrema in panel~(b). It also captures the relaxation towards zero after the final velocity change. No progressive phase shift or persistent force offset are observed. 

The agreement is slightly less uniform for the two histories with a net decrease in Reynolds number. In panel~(c), the main discrepancy appears near the final negative minimum and during the subsequent recovery. In panel~(d), the differences are concentrated immediately before and after the dominant positive maximum, particularly near the preceding negative minimum and during the following decay. These discrepancies should not be interpreted as evidence of a general directional bias, since the ramp direction, Reynolds number range, velocity variation and waveform all differ between the four cases. The different Reynolds number ranges are a possible contributor, particularly in panel~(c), but their effect cannot be isolated from the present comparisons.

The Mei-Adrian and Dorgan-Loth kernels again perform relatively well. Both capture the timing of most extrema and reproduce several parts of the response with reasonable accuracy. However, their agreement is less consistent near the largest extrema and during the final relaxation than that of the present kernel.

\section{Discussion and conclusions}

The aim of this work was to construct a history kernel for a spherical particle in a spatially uniform flow over the Reynolds-number range \(0.2\leq\Rey\leq200\) and to assess its predictive capability for diverse velocity histories. A new extraction methodology uses finite-duration ramp flows to reduce the history-force convolution to a finite-window integral of the kernel, facilitating its extraction. The approach combines in-ramp differentiation and a midpoint approximation of the post-ramp window average, using paired acceleration and deceleration simulations to obtain kernel values at individual Reynolds numbers. These are then assembled into a finite-\(\Rey\) kernel surface. To complete the surface where direct extraction is unavailable or unreliable, short-lag and low-\(\Rey\) treatments enforce the classical Basset limits, while a long-lag closure informed by the analyses of \citet{Mei_Adrian_1992} and \citet{Lawrence_Mei_1995} supplies the unresolved long-time decay.

Subsequently, the kernel was analysed in relation to wake dynamics. The non-monotonic behaviour observed at higher Reynolds numbers is associated with delayed boundary-layer and wake adjustment. The vorticity budgets revealed a greater relative importance of advection and stretching at higher Reynolds numbers. Decomposition of the kernel \(K\) showed that its pressure contribution increases sufficiently to outweigh the continuing decrease in its viscous contribution, giving rise to the observed non-monotonic behaviour. Finally the present kernel was evaluated for a range of velocity histories through comparisons with numerical simulations and other models available in the literature. The results show a substantial improvement over the finite-\(\Rey\) formulations considered, reproducing the main force extrema, phase, and relaxation across finite ramps, oscillatory forcing, and non-periodic composite histories without case-specific adjustment. More broadly, the proposed extraction methodology could be adapted to identify history contributions to hydrodynamic torque and spin-induced lift, and to investigate history forces in spatially non-uniform flows.

The validation cases, however, do not reveal a single threshold at which the performance of the present kernel begins to deteriorate. The observed discrepancies do not show a clear dependence on a single parameter. They cannot be attributed to acceleration alone, since some develop after the imposed acceleration has ended, whereas a high-frequency oscillation with a large acceleration amplitude is reconstructed accurately. The width of the sampled Reynolds-number range appears to influence the accuracy, but not independently of the forcing timescale. The short-lag treatment is unlikely to be the main source of error because the discrepancies are most apparent in later stages of finite-ramp validation, when the force is generated by older memory contributions and the short-lag part of the kernel no longer contributes. The low-\(\Rey\) closure is not active in the presented validation cases, while the prescribed long-lag treatment contributes less than \(1\,\%\) of the force in the decisive intervals and therefore cannot explain the main discrepancies. A moderate bias associated with the midpoint approximation or the core-tail stitching may contribute to discrepancies in individual cases. However, these approximations cannot explain the overall validation pattern, as the same regions of the kernel surface yield accurate predictions for many other histories. 

This validation pattern also raises the question of whether the dependence of \(K\) on \(s\) and \(\Rey(t-s)\) alone is sufficient to describe the memory contribution. This assumption may break down when large velocity changes substantially reorganise the boundary layer, shear layer and wake. The same values of \(s\) and \(\Rey(t-s)\) can correspond to different vorticity and pressure fields, depending on the velocity history before the source time and the subsequent evolution up to the observation time. A possible extension is to introduce a third state variable \(\Theta\)
describing the wake at the source time. The history force would then be written as
\[
F_{\mathrm H}(t)
=
\int_0^t
a_\infty(t-s)
\,
K\!\left(
s,
\Rey(t-s),
\Theta(t-s)
\right)
\,\mathrm ds .
\]
The present surface would correspond to the steady-wake condition
\[
\Theta=\Theta_{\mathrm{ss}}(\Rey),
\]
where \(\Theta_{\mathrm{ss}}\) denotes the value of the wake variable in the steady flow at a given Reynolds number. A minimal extension could represent the dependence on \(\Theta\) as
\[K(s,\Rey,\Theta)
\approx
K(s,\Rey)
+
\left[\Theta-\Theta_{\mathrm{ss}}(\Rey)\right]
\,
K_{\Theta}(s,\Rey),
\]
where \(K_{\Theta}\) describes how the memory kernel changes when the wake departs from its steady state. The variable \(\Theta\) should be identified from matched transient flows that reach the same Reynolds number after different preceding velocity histories. A suitable scalar wake measure would be one that assigns different values to these distinct flow states and reduces the remaining differences between their extracted responses. \\

\noindent\textbf{Acknowledgements.}\: 
 The authors acknowledge R. P. J. Kunnen for his input to this work and for valuable discussions. \\

\noindent\textbf{Author contributions.}\: 
MDL: Conceptualization, Methodology, Investigation, Validation, Visualization, Writing – original draft. JGMK: Conceptualization, Methodology, Supervision, Funding acquisition, Writing – review \& editing. SKFS: Conceptualization, Methodology, Supervision, Writing – review \& editing. \\

\noindent\textbf{Funding.}\:
This publication is part of the project “Particle and Fluid Dynamics in Magnetic Density Separation for Recycling of Plastic Waste” (project number 20833), which is (co-)funded by the Dutch Research Council (NWO). Computing time on Snellius, hosted by SURF, was provided by NWO through the Computing Time on National Computer Facilities call for proposals (project number 2025.012). \\

\noindent\textbf{Competing interests.}\:
The authors declare none. \\

\noindent\textbf{AI-use disclosure.}\:
During the preparation of this work, the authors used ChatGPT 5.5/5.6/6.0 between 04/2026 and 09/2026 to assist with code review, debugging and language refinement. The authors take full responsibility for the accuracy and integrity of the work. \\

\newpage
\appendix\label{appA}
\section{}\label{appA}
In this appendix, time, lag, velocity, acceleration and forces are dimensional unless explicitly normalised; kernel symbols denote the same dimensionless kernels evaluated using dimensional lag.
This appendix derives the high-\(\Rey\), long-lag closure introduced in \cref{sec:long_lag_treatment}. 

\subsection{Velocity step response}
\label{app:LM_local_kernel}
Consider an abrupt change in the imposed velocity from \(U_1\) to \(U_2\) at \(t=0\),
\begin{equation}
u_\infty(t)
=
U_1+(U_2-U_1)\,H(t),
\end{equation}
where \(H(t)\) is the Heaviside function. The corresponding acceleration is
\begin{equation}
\frac{\mathrm du_\infty}{\mathrm dt}
=
\Delta U\,\delta(t),
\qquad
\Delta U=U_2-U_1,
\end{equation}
with \(\delta(t)\) the Dirac delta function. For this finite velocity step, the history force may be written as
\begin{equation}
F_{\mathrm H}(t)
=
\Phi\,
\int_{-\infty}^{t}
K_{\mathrm{step}}(t-\tau;U_1,U_2)
\,
\frac{\mathrm du_\infty}{\mathrm d\tau}(\tau)
\,\mathrm d\tau .
\label{eq:appendix_step_convolution}
\end{equation}
Substitution of the impulsive acceleration gives
\begin{align}
F_{\mathrm H}(t)
&=
\Phi\,\Delta U\,
\int_{-\infty}^{t}
K_{\mathrm{step}}(t-\tau;U_1,U_2)\,
\delta(\tau)
\,\mathrm d\tau
\nonumber\\
&=
\Phi\,\Delta U\,
K_{\mathrm{step}}(t;U_1,U_2).
\label{eq:appendix_step_force_kernel}
\end{align}
For \(t>0\), the factor \(\delta(\tau)\) selects the source time \(\tau=0\), so the kernel is evaluated at the lag \(t-\tau=t\). Since the velocity step occurs at \(t=0\), the elapsed time coincides with the lag, \(s=t\). Therefore,
\begin{equation}
F_{\mathrm H}(s)
=
\Phi\,\Delta U\,
K_{\mathrm{step}}(s;U_1,U_2).
\label{eq:appendix_step_force_kernel_lag}
\end{equation}
Thus, following an ideal velocity step, the transient force and the corresponding step response kernel have the same dependence on elapsed time. They differ only by the constant factor \(\Phi\,\Delta U\). Equation \eqref{eq:appendix_step_force_kernel_lag} defines the effective response corresponding to the particular step \(U_1\rightarrow U_2\). The local kernel \(K(s,\Rey_0)\) used in the present surface is obtained below by taking the infinitesimal-step limit \(U_1\rightarrow U_2\).

\subsection{Infinitesimal-step limit of the Lawrence-Mei result}
\label{app:Infinitesimal-step limit}

Define the velocity ratio as 
\begin{equation}
\lambda
=
\frac{U_1}{U_2}
=
\frac{\Rey_1}{\Rey_2}.
\label{eq:appendix_velocity_ratio}
\end{equation}
The velocity change can then be written as
\begin{equation}
\Delta U
=
U_2\,(1-\lambda).
\label{eq:appendix_step_amplitude}
\end{equation}
Lawrence \& Mei normalise the transient force using the steady Stokes drag based on the final velocity,
\begin{equation}
\varphi_{\mathrm H}
=
\frac{F_{\mathrm H}}
     {6\,\upi\,\mu\,r\,U_2}
=
\frac{F_{\mathrm H}}
     {\Phi\,U_2}.
\label{eq:appendix_LM_force_normalization}
\end{equation}
Combining \cref{eq:appendix_step_force_kernel_lag} and
\cref{eq:appendix_step_amplitude} gives
\begin{equation}
\varphi_{\mathrm H}
=
(1-\lambda)\,
K_{\mathrm{step}}(s;U_1,U_2).
\label{eq:appendix_LM_step_kernel_relation}
\end{equation}
For a finite step between two positive velocities, \citet{Lawrence_Mei_1995} obtained the far-wake result
\begin{equation}
\varphi_{\mathrm H}
\sim
\frac{3}{\Rey_2}\,
\left(
\varphi_2+\Rey_2\,\varphi_2'
\right)
\,
\left(
\varphi_2-\varphi_1
\right)
\,
\xi^{-2},
\qquad
\xi
=
\frac{U_2\,s}{r}.
\label{eq:appendix_LM_finite_step}
\end{equation}
Here
\begin{equation}
\varphi_i
=
\varphi(\Rey_i),
\qquad
\varphi_i'
=
\left.
\frac{\mathrm d\varphi}{\mathrm d\Rey}
\right|_{\Rey_i},
\label{eq:appendix_drag_definitions}
\end{equation}
where
\begin{equation}
\varphi(\Rey)
=
\frac{F_{\mathrm{QS}}}
     {6\,\upi\,\mu\,r\,U}
=
\frac{F_{\mathrm{QS}}}
     {\Phi\,U}
\label{eq:appendix_steady_drag_correction}
\end{equation}
is the dimensionless steady-drag correction.

The factor \(\varphi_2-\varphi_1\) in \cref{eq:appendix_LM_finite_step} represents the difference between the volume-flux deficits carried by the old and new wakes. The factor \(\varphi_2+\Rey_2\,\varphi_2'\) describes the sensitivity of the complete steady drag to the final velocity.

At each Reynolds-number anchor, the intercept \(K(s,\Rey_0)\) represents the linearised response to an infinitesimal velocity perturbation about the corresponding steady base flow. The corresponding Lawrence-Mei kernel is obtained by taking an infinitesimal velocity step. Set
\begin{equation}
\Rey_2=\Rey,
\qquad
\Rey_1=\lambda\,\Rey,
\label{eq:appendix_local_limit_states}
\end{equation}
and let
\begin{equation}
\lambda\rightarrow1.
\end{equation}
We now express the same step-response kernel in terms of the dimensionless lag \(\xi=U_2\,s/r\) and the Reynolds numbers \(\Rey_i=U_i\,d/\nu_f\), writing it as \(K_{\mathrm{step}}(\xi;\Rey_1,\Rey_2)\).

From \cref{eq:appendix_LM_step_kernel_relation} and
\cref{eq:appendix_LM_finite_step},
\begin{equation}
K_{\mathrm{step}}(\xi;\lambda\,\Rey,\Rey)
\sim
\frac{3}{\Rey}\,
\left[
\varphi(\Rey)+\Rey\,\varphi'(\Rey)
\right]\,
\frac{
\varphi(\Rey)-\varphi(\lambda\,\Rey)
}{
1-\lambda
}
\,\xi^{-2}.
\label{eq:appendix_LM_kernel_before_limit}
\end{equation}
Taylor expansion of \(\varphi(\lambda\,\Rey)\) about \(\Rey\) gives
\begin{equation}
\varphi(\lambda\,\Rey)
=
\varphi(\Rey)
-
(1-\lambda)\,\Rey\,\varphi'(\Rey)
+
\mathcal{O}\left((1-\lambda)^2\right).
\label{eq:appendix_drag_taylor_expansion}
\end{equation}
Consequently,
\begin{equation}
\lim_{\lambda\rightarrow1}
\frac{
\varphi(\Rey)-\varphi(\lambda\,\Rey)
}{
1-\lambda
}
=
\Rey\,\varphi'(\Rey).
\label{eq:appendix_drag_limit}
\end{equation}
Substitution into \cref{eq:appendix_LM_kernel_before_limit} gives the adjusted far-wake kernel
\begin{equation}
K_{\mathrm{LM}}(\xi;\Rey)
\sim
3\,\varphi'(\Rey)\,
\left[
\varphi(\Rey)+\Rey\,\varphi'(\Rey)
\right]\,
\xi^{-2}.
\label{eq:appendix_LM_local_dimensionless}
\end{equation}
Returning to dimensional lag, \cref{eq:appendix_LM_local_dimensionless} becomes
\begin{equation}
K_{\mathrm{LM}}(s;\Rey)
\sim
\frac{A_{\mathrm{LM}}(\Rey)}{s^2},
\label{eq:appendix_LM_dimensional_kernel}
\end{equation}
where
\begin{equation}
A_{\mathrm{LM}}(\Rey)
=
3\,\varphi'(\Rey)\,
\left[
\varphi(\Rey)+\Rey\,\varphi'(\Rey)
\right]\,
\left(\frac{r}{U}\right)^2.
\label{eq:appendix_LM_dimensional_amplitude}
\end{equation}

\Cref{eq:appendix_LM_dimensional_kernel} establishes that the step-response kernel ultimately decays as \(s^{-2}\) and provides the corresponding formal Lawrence-Mei amplitude \(A_{\mathrm{LM}}(\Rey)\). The amplitude \(A_{\mathrm{LM}}(\Rey)\)  was evaluated using the numerical steady-drag curve. At each Reynolds number, the steady drag was obtained from the steady state solutions and normalised according to \cref{eq:appendix_steady_drag_correction}. Subsequently, \(\varphi'(\Rey)\) was evaluated using finite differences.  Substitution into \cref{eq:appendix_LM_dimensional_amplitude} provides an estimate of the ultimate inverse-square amplitude.

\subsection{Empirical transition to the inverse-square limit}
\label{app:effective_amplitude_fitting}

The Lawrence-Mei result implies that, for positive non-reversing motion, the kernel approaches an inverse-square decay at sufficiently large lag,
\begin{equation}
K(s,\Rey)
\sim
A_\infty(\Rey)\,s^{-2},
\qquad
s\rightarrow\infty,
\label{eq:appendix_far_wake_limit}
\end{equation}
where \(A_\infty(\Rey)\) is the far-wake amplitude. It follows that the quantity
\begin{equation}
A_{\mathrm{eff}}(s,\Rey)
=
s^2\,K(s,\Rey)
\label{eq:appendix_effective_amplitude}
\end{equation}
approaches a constant in the far-wake regime,
\begin{equation}
A_{\mathrm{eff}}(s,\Rey)
\rightarrow
A_\infty(\Rey)
\qquad
\text{as } s\rightarrow\infty .
\end{equation}
At finite lag, however, \(A_{\mathrm{eff}}\) does not need to be constant. Its variation therefore provides a measure of how the kernel approaches the ultimate inverse-square regime.

At the measured long-lag handoff,
\begin{equation}
K_{\mathrm{end}}(\Rey)
=
K_{\mathrm{surf}}(s_{\mathrm{end}},\Rey),
\end{equation}
and
\begin{equation}
A_{\mathrm{end}}(\Rey)
=
s_{\mathrm{end}}^2\,K_{\mathrm{end}}(\Rey).
\label{eq:appendix_handoff_amplitude}
\end{equation}
The ratio between the ultimate and handoff amplitudes is
\begin{equation}
r_\infty(\Rey)
=
\frac{A_\infty(\Rey)}
     {A_{\mathrm{end}}(\Rey)}
=
\lim_{s\rightarrow\infty}
\frac{s^2\,K(s,\Rey)}
     {s_{\mathrm{end}}^2\,K_{\mathrm{end}}(\Rey)}.
\label{eq:appendix_rinf_definition}
\end{equation}
The evaluated ratios \(A_{\mathrm{LM}}(\Rey)/A_{\mathrm{end}}(\Rey)\) are close to \(0.08\) for the examined high-\(\Rey\) cases. The closure therefore adopts the common value \(r_\infty=0.08\), giving \(A_\infty(\Rey)=0.08\,A_{\mathrm{end}}(\Rey)\) as an approximation to the Lawrence-Mei amplitude \(A_{\mathrm{LM}}(\Rey)\).

The long-duration intercepts indicate that \(A_{\mathrm{eff}}\) relaxes from \(A_{\mathrm{end}}\) towards a substantially smaller limiting value. The closure assumes an exponential relaxation of \(A_{\mathrm{eff}}\) with the distance convected by the wake, consistent with the approximately exponential decay observed in the long-duration records. This approach is represented empirically by
\begin{equation}
\frac{
A_{\mathrm{eff}}(s,\Rey)-A_\infty(\Rey)
}{
A_{\mathrm{end}}(\Rey)-A_\infty(\Rey)
}
=
\exp\left(-\frac{\eta}{L_\eta(\Rey)}\right),
\qquad
\eta
=
\frac{U\,(s-s_{\mathrm{end}})}{d}.
\label{eq:appendix_amplitude_defect}
\end{equation}
The coordinate \(\eta\) is the distance travelled by the wake after the handoff divided by the particle diameter. The parameter \(L_\eta\) is therefore a relaxation distance measured in particle diameters. Using \cref{eq:appendix_rinf_definition}, \Cref{eq:appendix_amplitude_defect} becomes
\begin{equation}
\frac{A_{\mathrm{eff}}(s,\Rey)}
     {A_{\mathrm{end}}(\Rey)}
=
r_\infty(\Rey)
+
\left(1-r_\infty(\Rey)\right)\,
\exp\left(-\frac{\eta}{L_\eta(\Rey)}\right).
\label{eq:appendix_normalised_amplitude}
\end{equation}
Substitution of \cref{eq:appendix_effective_amplitude} and \cref{eq:appendix_handoff_amplitude} gives the dimensional high-\(\Rey\), long-lag kernel
\begin{equation}
K_{\mathrm{HL}}(s,\Rey)
=
K_{\mathrm{end}}(\Rey)\,
\left(\frac{s_{\mathrm{end}}}{s}\right)^2\,
\left[
r_\infty(\Rey)
+
\left(1-r_\infty(\Rey)\right)\,
\exp\left(-\frac{\eta}{L_\eta(\Rey)}\right)
\right].
\label{eq:appendix_high_re_kernel}
\end{equation}
At the handoff, \(\eta=0\), and therefore
\begin{equation}
K_{\mathrm{HL}}(s_{\mathrm{end}},\Rey)
=
K_{\mathrm{end}}(\Rey).
\end{equation}
The closure is continuous in value, while continuity of its lag derivative is not imposed.
Finally, dividing by \(K_{\mathrm B}(s)\) gives the Basset-normalised long-lag closure:
\begin{equation}
R_{\mathrm{HL}}(s,\Rey)
=
R_{\mathrm{end}}(\Rey)\,
\left(\frac{s_{\mathrm{end}}}{s}\right)^{3/2}\,
\left[
r_\infty(\Rey)
+
\left(1-r_\infty(\Rey)\right)\,
\exp\left(-\frac{\eta}{L_\eta(\Rey)}\right)
\right].
\label{eq:appendix_high_re_ratio}
\end{equation}

\bibliographystyle{jfm}
\bibliography{jfm}

@InBook{Basset1888,
  author    = {Basset, A. B.},
  title     = {A Treatise on Hydrodynamics: With Numerous Examples},
  year      = {1888},
  publisher = {Deighton, Bell and Co.},
  address   = {Cambridge},
}

@article{MR1983,
  author    = {Maxey, M. R. and Riley, J. J.},
  title     = {Equation of motion for a small rigid sphere in a nonuniform flow},
  journal   = {Phys. Fluids},
  year      = {1983},
  volume    = {26},
  number    = {4},
  pages     = {883--889},
  doi       = {10.1063/1.864230},
  url       = {https://doi.org/10.1063/1.864230},
  issn      = {0031-9171},
}

@article{BOUSSINESQ1885,
  author    = {Boussinesq, J.},
  title     = {Sur la r{\'e}sistance qu'oppose un liquide ind{\'e}fini en repos, sans pesanteur, au mouvement vari{\'e} d'une sph{\`e}re solide qu'il mouille sur toute sa surface, quand les vitesses restent bien continues et assez faibles pour que leurs carr{\'e}s et produits soient n{\'e}gligeables},
  journal   = {C. R. Acad. Sci. Paris},
  year      = {1885},
  volume    = {100},
  pages     = {935--937},
  url       = {https://cir.nii.ac.jp/crid/1571135650072086912},
}

@article{Odar_Hamilton_1964,
  author    = {Odar, F. and Hamilton, W. S.},
  title     = {Forces on a sphere accelerating in a viscous fluid},
  journal   = {J. Fluid Mech.},
  year      = {1964},
  volume    = {18},
  number    = {2},
  pages     = {302--314},
  doi       = {10.1017/S0022112064000210},
  url       = {https://doi.org/10.1017/S0022112064000210},
}

@article{Mei_Adrian_1992,
  author    = {Mei, R. and Adrian, R. J.},
  title     = {Flow past a sphere with an oscillation in the free-stream velocity and unsteady drag at finite {Reynolds} number},
  journal   = {J. Fluid Mech.},
  year      = {1992},
  volume    = {237},
  pages     = {323--341},
  doi       = {10.1017/S0022112092003434},
  url       = {https://doi.org/10.1017/S0022112092003434},
}

@inbook{Batchelor_1967,
  author    = {Batchelor, G. K.},
  title     = {An Introduction to Fluid Dynamics},
  year      = {1967},
  pages     = {453},
  publisher = {Cambridge University Press},
}

@article{Chang_Maxey_1995,
  author    = {Chang, E. J. and Maxey, M. R.},
  title     = {Unsteady flow about a sphere at low to moderate {Reynolds} number. {Part} 2. {Accelerated} motion},
  journal   = {J. Fluid Mech.},
  year      = {1995},
  volume    = {303},
  pages     = {133--153},
  doi       = {10.1017/S0022112095004204},
  url       = {https://doi.org/10.1017/S0022112095004204},
}

@article{Sano_1981,
  author    = {Sano, T.},
  title     = {Unsteady flow past a sphere at low {Reynolds} number},
  journal   = {J. Fluid Mech.},
  year      = {1981},
  volume    = {112},
  pages     = {433--441},
  doi       = {10.1017/S0022112081000499},
  url       = {https://doi.org/10.1017/S0022112081000499},
}

@article{Bentwich_Miloh_1978,
  author    = {Bentwich, M. and Miloh, T.},
  title     = {The unsteady matched {Stokes}--{Oseen} solution for the flow past a sphere},
  journal   = {J. Fluid Mech.},
  year      = {1978},
  volume    = {88},
  number    = {1},
  pages     = {17--32},
  doi       = {10.1017/S0022112078001962},
  url       = {https://doi.org/10.1017/S0022112078001962},
}

@article{Dennis_Walker_1972,
  author    = {Dennis, S. C. R. and Walker, J. D. A.},
  title     = {Numerical solutions for time-dependent flow past an impulsively started sphere},
  journal   = {Phys. Fluids},
  year      = {1972},
  volume    = {15},
  number    = {4},
  pages     = {517--525},
  doi       = {10.1063/1.1693943},
  url       = {https://doi.org/10.1063/1.1693943},
  issn      = {0031-9171},
}

@article{Mei_Lawrence_Adrian_1991,
  author    = {Mei, R. and Lawrence, C. J. and Adrian, R. J.},
  title     = {Unsteady drag on a sphere at finite {Reynolds} number with small fluctuations in the free-stream velocity},
  journal   = {J. Fluid Mech.},
  year      = {1991},
  volume    = {233},
  pages     = {613--631},
  doi       = {10.1017/S0022112091000629},
  url       = {https://doi.org/10.1017/S0022112091000629},
}

@article{Mei_1993,
  author    = {Mei, R.},
  title     = {History force on a sphere due to a step change in the free-stream velocity},
  journal   = {Intl J. Multiphase Flow},
  year      = {1993},
  volume    = {19},
  number    = {3},
  pages     = {509--525},
  doi       = {10.1016/0301-9322(93)90064-2},
  url       = {https://doi.org/10.1016/0301-9322(93)90064-2},
  issn      = {0301-9322},
}

@article{Mei_1994,
  author    = {Mei, R.},
  title     = {Flow due to an oscillating sphere and an expression for unsteady drag on the sphere at finite {Reynolds} number},
  journal   = {J. Fluid Mech.},
  year      = {1994},
  volume    = {270},
  pages     = {133--174},
  doi       = {10.1017/S0022112094004222},
  url       = {https://doi.org/10.1017/S0022112094004222},
}

@article{Chang_Maxey_1994,
  author    = {Chang, E. J. and Maxey, M. R.},
  title     = {Unsteady flow about a sphere at low to moderate {Reynolds} number. {Part} 1. {Oscillatory} motion},
  journal   = {J. Fluid Mech.},
  year      = {1994},
  volume    = {277},
  pages     = {347--379},
  doi       = {10.1017/S002211209400279X},
  url       = {https://doi.org/10.1017/S002211209400279X},
}

@article{Lawrence_Mei_1995,
  author    = {Lawrence, C. J. and Mei, R.},
  title     = {Long-time behaviour of the drag on a body in impulsive motion},
  journal   = {J. Fluid Mech.},
  year      = {1995},
  volume    = {283},
  pages     = {307--327},
  doi       = {10.1017/S0022112095002333},
  url       = {https://doi.org/10.1017/S0022112095002333},
}

@article{Lovalenti_Brady_1995,
  author    = {Lovalenti, P. M. and Brady, J. F.},
  title     = {The temporal behaviour of the hydrodynamic force on a body in response to an abrupt change in velocity at small but finite {Reynolds} number},
  journal   = {J. Fluid Mech.},
  year      = {1995},
  volume    = {293},
  pages     = {35--46},
  doi       = {10.1017/S0022112095001625},
  url       = {https://doi.org/10.1017/S0022112095001625},
}

@article{KIM_ELGHOBASHI_1998,
  author    = {Kim, I. and Elghobashi, S. and Sirignano, W. A.},
  title     = {On the equation for spherical-particle motion: effect of {Reynolds} and acceleration numbers},
  journal   = {J. Fluid Mech.},
  year      = {1998},
  volume    = {367},
  pages     = {221--253},
  doi       = {10.1017/S0022112098001657},
  url       = {https://doi.org/10.1017/S0022112098001657},
}

@article{WAKABA_2005,
  author    = {Wakaba, L. and Balachandar, S.},
  title     = {History force on a sphere in a weak linear shear flow},
  journal   = {Intl J. Multiphase Flow},
  year      = {2005},
  volume    = {31},
  number    = {9},
  pages     = {996--1014},
  doi       = {10.1016/j.ijmultiphaseflow.2005.05.009},
  url       = {https://doi.org/10.1016/j.ijmultiphaseflow.2005.05.009},
  issn      = {0301-9322},
}

@article{DORGAN2007,
  author    = {Dorgan, A. J. and Loth, E.},
  title     = {Efficient calculation of the history force at finite {Reynolds} numbers},
  journal   = {Intl J. Multiphase Flow},
  year      = {2007},
  volume    = {33},
  number    = {8},
  pages     = {833--848},
  doi       = {10.1016/j.ijmultiphaseflow.2007.02.005},
  url       = {https://doi.org/10.1016/j.ijmultiphaseflow.2007.02.005},
  issn      = {0301-9322},
}

@article{Ern2012WakeInducedOP,
  author    = {Ern, P. and Risso, F. and Fabre, D. and Magnaudet, J.},
  title     = {Wake-induced oscillatory paths of bodies freely rising or falling in fluids},
  journal   = {Annu. Rev. Fluid Mech.},
  year      = {2012},
  volume    = {44},
  pages     = {97--121},
  doi       = {10.1146/annurev-fluid-120710-101250},
  url       = {https://doi.org/10.1146/annurev-fluid-120710-101250},
}

@book{happel1983low,
  author    = {Happel, J. and Brenner, H.},
  title     = {Low Reynolds Number Hydrodynamics: With Special Applications to Particulate Media},
  year      = {1983},
  volume    = {1},
  publisher = {Martinus Nijhoff Publishers},
  address   = {The Hague},
  series    = {Mechanics of Fluids and Transport Processes},
  doi       = {10.1007/978-94-009-8352-6},
  url       = {https://doi.org/10.1007/978-94-009-8352-6},
  isbn      = {9789024728770},
}

@phdthesis{Moorman1955,
  author    = {Moorman, R. W.},
  title     = {Motion of a spherical particle in the accelerated portion of free fall},
  year      = {1955},
  address   = {Iowa},
  school    = {University of Iowa},
}

\end{document}